\documentclass{article}
 \usepackage{amssymb,amsfonts,amsmath}

\newcommand{\<}{\langle}
\renewcommand{\>}{\rangle}
\newcommand{\R}{\mathcal{R}}
\newcommand{\Sc}{\mathcal{S}}
\newcommand{\Lc}{\mathcal{L}}
\newcommand{\E}{\mathcal{E}}
\newcommand{\F}{\mathcal{F}}
\newcommand{\A}{\mathcal{A}}
\newcommand{\Hc}{\mathcal{H}}
\newcommand{\K}{\mathcal{K}}
\newcommand{\D}{\mathcal{D}}
\newcommand{\J}{\mathcal{J}}
\newcommand{\N}{\mathcal{N}}
\newcommand{\M}{\mathcal{M}}
\newcommand{\Oc}{\mathcal{O}}

\newcommand{\mR}{\mathbb{R}}
\newcommand{\mN}{\mathbb{N}}

\newcommand{\1}{{\bf 1}}
\newcommand{\dsp}{\displaystyle}
\newcommand{\txt}{\textstyle}
\newcommand{\df}{:=}
\newcommand{\con}{\mathrm{const.}}

\newcommand{\FF}{{\Phi_0}}
\newcommand{\WF}{{W_0}}
\newcommand{\aq}{\vec{q}}
\newcommand{\ap}{\vec{p}}
\newcommand{\ar}{\vec{r}}
\newcommand{\as}{\vec{s}}
\newcommand{\ak}{\vec{\xi}}
\newcommand{\aet}{\vec{\eta}}
\newcommand{\ax}{\vec{x}}
\newcommand{\ay}{\vec{y}}
\newcommand{\az}{\vec{z}}
\newcommand{\ep}{\epsilon}

\newcommand{\n}{\nabla}

\newcommand{\w}{\omega}
\newcommand{\W}{\Omega}

\newcommand{\ov}{\overline}
\newcommand{\wh}{\widehat}

\DeclareMathOperator{\Ker}{Ker} \DeclareMathOperator{\Ran}{Ran}
\DeclareMathOperator{\Ip}{Im} \DeclareMathOperator{\Rp}{Re}
\DeclareMathOperator{\id}{id} \DeclareMathOperator{\Tr}{Tr}

\title{Quantum backreaction (Casimir) effect\\I. What are admissible idealizations?}
\author{Andrzej Herdegen%
\thanks{e-mail: herdegen@th.if.uj.edu.pl}\\
{\it Institute of Physics, Jagiellonian University,}\\
{\it Reymonta 4, 30-059 Cracow, Poland}}
\date{}
\begin{document}

\maketitle

\begin{abstract}
  Casimir effect, in a~broad interpretation which we adopt here,
  consists in a~backreaction of a~quantum system to adiabatically
  changing external conditions. Although the system is usually
  taken to be a~quantum field, we show that this restriction
  rather blurs than helps to clarify the statement of the problem.
  We~discuss the problem from the point of view of algebraic
  structure of quantum theory, which is most appropriate in this
  context. The system in question may be any quantum system, among others
  both finite as infinite dimensional canonical systems are
  allowed. A~simple finite-dimensional model is discussed.
  We~identify precisely the source of difficulties and infinities in
  most of traditional treatments of the problem for infinite
  dimensional systems (such as quantum fields), which is
  incompatibility of algebras of observables or their
  representations. We~formulate
  conditions on model idealizations which are acceptable for the
  discussion of the adiabatic backreaction problem. In~the case of
  quantum field models in that class we find that the normal
  ordered energy density is a~well defined distribution,
  yielding global energy in the limit of a~unit test function.
  Although we see the
  ``zero point'' expressions as inappropriate, we show how they can
  arise in the quantum field theory context as a~result of
  uncontrollable manipulations.

\vspace*{2ex} \noindent PACS numbers: 03.70.+k, 03.65.Bz, 11.10.-z
\end{abstract}
\vfill \eject

\renewcommand{\theequation}{\thesection.\arabic{equation}}

\section{Introduction}\label{int}

The Casimir effect, bearing its name from the pioneer work by
Casimir~\cite{cas},  has become in recent decades an increasingly
popular topic in quantum field theory, with a~new review of the
subject appearing every few years, see~\cite{pmg,er,wm,kam,bmm}.
The effect consists in the response of a~quantum field, even in
a~ground state, to the introduction of external, usually
macroscopic, bodies. Initially the effect existed as a~theoretical
prediction only, and a~rather mysterious one, for that matter.
However, increasing experimental evidence of its existence (see
e.g.~\cite{bmm}) has lead to attempts at better understanding of
its theoretical foundation. The problem is, that the theoretical
side of the phenomenon has been plagued from the beginning by
divergent expressions, as well as conceptual difficulties, which
have proved to be surprisingly persistent. This is the more
surprising, that models usually considered in this context are
\emph{linear}, so the usual sources of quantum field infinities
are absent here.

In~an earlier paper~\cite{her} I have given a~diagnosis of the
reasons of this state of affairs and proposed to treat the problem
from the algebraic point of view. This is the most natural and
fruitful framework in quantum physics, with its beginnings already
in the classical book on quantum mechanics by Dirac, and modern
developments in quantum field theory and statistical physics
described e.g.\ in monographs~\cite{haa} and~\cite{br}. When
viewed from that angle the source of difficulties is rather
obvious, and can be briefly termed as uncritical use of the
concept of quantum field~\cite{her}. More precisely, what we mean
is this. The first step to define a~quantum theory is to identify
a set of quantum observables (we ignore here the question of
non-observable variables) together with algebraic relations
between them, such as canonical commutation relations. Once we
have this, a~concrete physical realization of the theory
corresponds to a~choice of a~representation of the algebra of
observables. Non-comparable physical situations are realized by
non-equivalent representations~\cite{haa}. Although we want to see
the real world as a~unity, physics, of course, is about
idealizations, and various idealizations need not be compatible
(take e.g.\ an isolated system and a~thermodynamic limit system).
However, if we want to consider transitions from one physical
situation to another and compare values of one and the same
observable in various states, all situations taken into account
must be describable in one common representation. Now, these
scheme is violated in most treatments of the Casimir effect. For
a~typical situation of a~quantum field in a~region with movable
sharp boundaries the difficulty arises already on the algebraic
level: there is no consistent choice of an algebra of observables
for all physical situations coming into play. The energy of the
``free'' field is an observable defined in the vacuum
representation of the algebra of the field smeared with Schwartz
test functions. For this algebra evolutions imposed on the field
by the presence of boundaries cannot be defined. Furthermore, even
if one ``smooths out'' the boundaries so as to make a~common
choice of an algebra possible, one still has to satisfy rather
severe restrictions necessary to ensure the equivalence of
representations. These restrictions are typically violated in
usual treatments. For these reasons we have advocated in
\cite{her} the view, that the model of sharp boundaries, as well
as many other insufficiently regular models, are wrong
idealizations in the context of Casimir effect, and we have also
proposed (and analyzed) a~class of models imitating Dirichlet
conditions. Let us stress this: once a~model has correctly been
chosen, there is no space (nor need) for further \emph{ad hoc}
regularizations, and the formalism should yield well-defined
answers to legitimate questions. Although views on nonphysical
nature of sharp boundary conditions have been also expressed
elsewhere (see e.g.~\cite{dc}), it seems that the conditions for
a~model to be acceptable in the sense described above have not
been analyzed before. For instance, in a~series of recent papers
Graham \emph{et al.}~\cite{gj} investigate a~\emph{linear} model
imitating Dirichlet conditions. Being linear, the model should be
well-defined without any renormalization (except for a~trivial
normal ordering for quadratic quantities like energy density; for
external potentials without bounded states this is our example
(iii)$_1$ at the end of Section~\ref{cq} below). However,
renormalization is \emph{ad hoc} imposed on it by the authors in
order to give meaning to a~meaningless expression.

The algebraic problems we have described often do not appear if
one restricts attention to \emph{local} quantities in quantum
field theory. This fact is connected with what in algebraic
formulation is called the local quasiequivalence of
representations (see~\cite{haa}). The point is as follows.
In~quantum field theory observables are equipped with the property
of locality: each local observable carries as a~label an open
spacetime region with compact closure in which it may be measured.
As~stated above, two representations of the totality of these
observables representing two different physical situations may be
non-equivalent. However, physically one would expect that even if
the two situations are globally non-comparable, one should be able
to compare results of \emph{local} measurements (think of the
vacuum representation and a~thermodynamic limit representation).
Mathematical formulation of this expectation is this: think of
states in each of the representations as density operators;
restrict attention to an arbitrarily chosen compact region of
spacetime; then for each state in one of the representations there
is a~state in the other which yields the same expectation values
for observables localized in the chosen region. If the two
representations have this property they are called locally
quasiequivalent. As~it turns out physically important
representations do indeed often have this relative property, and
then expectation values of local quantities may be compared.
However, in the situation we want to consider in this paper this
result falls short of our needs in twofold way. First, we want to
calculate expectation values of global quantities, which are
limits of local ones for the size of the spacetime region tending
to infinity -- in this case the global differences of the
representations come into play. Second, in situations like fields
with imposed boundary conditions, even finite regions which
overlap with boundaries are not local in the above sense: for
those regions even the scopes of local algebras in presence of
boundaries are different than in the vacuum theory.

Another important point we want to stress in our analysis of
Casimir effect is the choice of the observable to be compared in
various considered states. In~our view the backreaction of
a~system perturbed by external agents is determined by the
expectation value of the energy as defined by the unperturbed
system, one and the same (as an operator) in all states to be
considered. A~more systematic discussion of this point in a~wider
context will be found in the next section. Here we want to note
that some \emph{local}, in the spirit of the last paragraph,
calculations of the Casimir energy do follow similar ideology; in
the gravitational context see esp.\ a~paper by Kay~\cite{kay}, and
for electromagnetic field with conducting boundaries a~paper by
Scharf and Wreszinski~\cite{sw}. However, in many other local
calculations, esp.\ those using ``the Green function method'', the
situation is somewhat ambiguous: it is often not clear enough what
the general viewpoint is, and the result may agree with the above
method in some cases, but disagree in others. We~shall discuss
this point more fully in Section~\ref{loc} below. For the
\emph{global} energy, as determined by the unperturbed system, to
be defined in states of the system influenced by external
conditions, as required by the above ideology, we need one common
algebra and globally equivalent representations, as explained
earlier. This imposes restrictions on the perturbed dynamics,
which are usually violated, and the transition from local Casimir
energy to global one is then blocked by infinities of
\emph{physical} nature. Any ``regularization'' thereof is an ad
hoc procedure, striving at this late stage to compensate for the
wrong idealization in interaction with external conditions.
Finally, there is a~group of works explicitly comparing the
expectation values of different global observables: energy with
and without interaction. Here, apparently, is the place of the
``zero point'' ideology. We~shall come back to this point later
on, here we only note, that in this case infinities are even more
likely to appear. In~that method one subtracts expectation values
of regularized ``bare'' energy observables; different energy
observables may have different singularities, not cancelling under
subtraction.

The present paper is the first of the two in which we develop and
describe more fully what was announced in~\cite{her} (we use
notation slightly changed at some points with respect to that
paper). Here we discuss more general results on the admissibility
of models for the purpose of investigation of quantum
backreaction. In~the second paper applications to particular
models are discussed. We~use rigorous mathematics, and present
real proofs. However, we hope that the paper is readable for
a~wide audience.

In~Section~\ref{ext} we place the quantum field Casimir effects in
a wider context of a~backreaction of a~quantum system to adiabatic
changes. This section thoroughly discusses the foundation for the
calculation of this backreaction in any quantum system. In~Section
\ref{cq} we discuss quantization of a~class of linear systems,
which include quantum fields under linear external perturbations.
We~put stress on less widely known aspects of this otherwise
standard procedure which are important in the present context.
Section~\ref{fd} discusses an application to a~finite-dimensional
system. In~Section~\ref{id} we treat infinite-dimensional cases,
and we formulate conditions for admissibility of a~model for the
discussion of backreaction effects. More specifically, we consider
a quantum field case in Section~\ref{loc}. We~show that with
a~slight strengthening of these conditions not only global energy,
but also energy density may be defined, and in the appropriate
limit global energy is recovered. Section~\ref{zem} contains
somewhat more explicit discussion of the points made earlier in
this Introduction on the existing calculations of Casimir effect.
We~also comment there on the ``zero point'' expressions for the
Casimir energy. We~try to understand, from the point of view of
the formalism presented in the present work, how such expressions
may arise. We~show how imposing unacceptable idealization of sharp
boundaries and doing unjustified manipulations leads from our
expression for the energy density to ``zero point'' expressions
for Casimir energy. Appendix gives a~simple form to a~handful of
mathematical facts in Fock space which are needed in the main
text. These are known results, but we believe that this summary
makes some of them more accessible.

\setcounter{equation}{0}
\section{A~quantum system under external conditions}\label{ext}

Trying to put the discussed phenomenon in a~broader context we
shall adopt the following point of view. The Casimir-type effect
consists in the backreaction of a~quantum system on the
adiabatically changing external conditions under which the system
is placed.

The background for this idea is this. We~consider a~larger closed
system consisting of two subsystems $Q$ and $M$. These subsystems
interact with each other, but to certain degree (this will be made
more precise below) maintain their separate identity. Part $Q$ is
our relatively simple quantum system under consideration (say,
electromagnetic field), while $M$ is supposed to be of much more
complicated nature (say, conductor plates), and to have among its
variables some of collective, macroscopic type (separation of the
plates). We~want to determine the effect of the evolution of the
joint system on the collective variables attached to $M$.

Because of the complicated nature of the part $M$ of the system
and its interaction with $Q$, to tackle the problem one has to
make some simplifying assumptions. There are at least two
possibilities, both of them of phenomenological nature. In~both
cases one simply represents part $M$ of the system by a~few
collective variables (such as separation of the plates),
suppressing all the details of this subsystem, and representing
the interaction between $M$ and $Q$ by some simple effective
model. The first possibility is to equip the collective variables
with a~fully quantum nature, and put forward a~simple model for
the closed system. This approach, when applied to the more
specific situation of a~quantum field in interaction with
macroscopic bodies, is chronologically more recent one in this
field, and is called the dynamic Casimir effect (see~\cite{bmm}).
Although we admit that this forms an open possibility, we shall
not take it up in this article. Firstly, not much can be said with
high degree of certainty and mathematical rigor. Secondly, the
apparent attractiveness of the approach does not necessarily
withstand a~closer scrutiny. A~macroscopic body undergoes
``constant observation'', so effects of decoherence play primary
role, which is not taken into account in this approach.
\pagebreak[1]

Another possibility, which we take up in this paper, has more
restricted aspirations, but admits mathematically rigorous
results, as we are going to argue below. We~have to admit,
however, that there is some confusion at its physical formulation.
We~hope to contribute to its removal. This second approach
consists in approximating the collective quantities, which
characterize a~macroscopic body as a~whole, by classical
variables. Moreover, one considers only situations, in which the
whole system changes adiabatically. The effect of the evolution on
the macroscopic (classical) variables in this context is what we
referred to as a~Casimir-type effect at the beginning of this
section. More specifically, the Casimir effect refers to a~quantum
field in interaction with macroscopic bodies.

One should be more specific about physical assumptions and
approximations involved in the situation implied in the last
paragraph. This is, in our opinion, a~point not clear enough in
many discussions of the Casimir effect. Therefore we shall try to
be systematic, even at a~risk of being too detailed.
\begin{itemize}
\item[(i)] One considers first the isolated quantum system $Q$
($M$ is absent). We~give its description in the algebraic
formulation of the Heisenberg picture, see e.g.~\cite{br}.
 \begin{itemize}
 \item[(i)$_1$] Basic quantum variables at a~fixed time form an abstract
 *-algebra $\mathcal{A}$, e.g.\ an algebra of canonical commutation relations
 (or, more technically, its exponentiation to the Weyl form).
 \item[(i)$_2$] This algebra is represented by operators in a~Hilbert
 space $\mathcal{H}$:
 \begin{equation}\label{ext-rep}
 \begin{split}
 \pi:\mathcal{A}\mapsto\pi(\mathcal{A})\,,\quad
 A\mapsto\pi(A)\,,&\quad \pi(A^*)=\pi(A)^*\,,\\
 \pi(\alpha A+\beta B)=\alpha\pi(A)+\beta\pi(B)\,,&\quad
 \pi(AB)=\pi(A)\pi(B)\,,
 \end{split}
 \end{equation}
 where $\pi(\mathcal{A})$ is a~concrete algebra of operators in
 $\mathcal{H}$. Vectors in that space, or, more generally, density
 operators acting in this space, represent states of the system $Q$.
 Representation $\pi$ is assumed to be irreducible; then vectors
 correspond to \emph{pure} states.
 \item[(i)$_3$] The intrinsic dynamics of $Q$ is defined by an
 automorphism of the algebra~$\mathcal{A}$:
 \begin{equation}\label{ext-ev0}
 \alpha_t:\mathcal{A}\mapsto\mathcal{A}\,,\quad
 A\mapsto\alpha_tA\,.
 \end{equation}
 This automorphism is implemented by a~unitary evolution in the
 Hilbert space $\mathcal{H}$:
 \begin{equation}\label{ext-uev0}
 \pi(\alpha_tA)=U(t)\pi(A)U(t)^*\,,\quad
 U(t)=\exp(itH)\,,
 \end{equation}
 where $H$ has the interpretation of the energy operator of the
 system. This operator is supposed to have nonnegative spectrum, and
 usually is assumed to have a~ground state, represented by a~unit
 eigenvector to the lowest point in the spectrum.
 One does not perturb the above
 relations by adding a~multiple of the identity operator to $H$, so
 the ground state may be assumed to have zero energy. By
 irreducibility of~$\pi$ the energy operator $H$ is then uniquely
 determined.
 \end{itemize}

\item[(ii)] One introduces now part $M$ into the system. This part
is characterized by classical variables (we shall denote them by
$a$), so no new quantum variables are added. Therefore system~$Q$
should retain its identity, and changes in its state will
influence the classical variables of $M$. Thus various states to
be considered must be physically comparable. These assumptions
have mathematical consequences.
 \begin{itemize}
 \item[(ii)$_1$] Identity of the system $Q$ is formed by the algebra
 $\mathcal{A}$ ((i)$_1$ above), so this algebra must remain unaffected by $M$.
 \item[(ii)$_2$] Physical comparability of states demands that also the
 particular representation $\pi$ of $\mathcal{A}$ ((i)$_2$ above)
 remains unaffected by the introduction of~$M$.
 \end{itemize}
 We~stress the importance of this point as
 it is both crucial for the scheme, as we see it, and usually
 overlooked. If various physical situations to be considered demanded
 different algebras or different (nonequivalent) representations,
 the approximation would break down, as one could not follow the
 change in the system $Q$ brought about by the creation of (and
 changes in) $M$, and its reaction to that occurrence. Further
 support for this point will be found below. Let us note
 again, what was discussed in introduction, that the local
 quasiequivalence of representations if $Q$ is a~quantum field
 system is not enough for our purposes.

\item[(iii)] We~consider now dynamics in presence of $M$, and
assume at first that the variables $a$ are frozen. In~this case
$Q$ is still a~closed system in interaction with conditions
created by $M$, and for each fixed $a$ its evolution is again
given by an automorphism of the algebra $\mathcal{A}$:
\begin{equation}\label{ext-eva}
 \alpha_{at}:\mathcal{A}\mapsto\mathcal{A}\,,\quad
 A\mapsto\alpha_{at}A\,.
\end{equation}
 One assumes implementability of new evolutions in the representation
 $\pi$: for each $a$ we have
\begin{equation}\label{ext-ueva}
 \pi(\alpha_{at}A)=U_a(t)\pi(A)U_a(t)^*\,,\quad
 U_a(t)=\exp(itH_a)\,.
\end{equation}
For each $a$ the generator $H_a$ is defined by this up to the
addition of a multiple of the identity operator, so we have the
freedom
\begin{equation}\label{ext-gau}
 H_a \rightarrow H_a+\lambda_a\id\,,
\end{equation}
where $\lambda_a$ is any real function of parameters $a$.

\item [(iv)] One allows now the coupled system $Q\!-\!M$ to
evolve. Part $Q$ alone is not a~closed system any more, so it
could be too restrictive to assume that the evolution of its
variables would be given at the algebraic level, as an
automorphism. However, this evolution should still be describable
in terms of unitary operators in the Hilbert space $\mathcal{H}$
(not forming a~one-parameter group, in general); this corresponds
to the assumption of conservation of~probabilities in the
subsystem $Q$. The use of the Schr\"odinger picture for the
quantum part $Q$ will be more convenient in the present context.
State of the coupled system $Q\!-\!M$ is specified at a~given time
by a~vector in the Hilbert space $\mathcal{H}$ (describing the
state of $Q$), and values of $a$ and, possibly, their time
derivatives. We~formulate the evolution of this system.
 \begin{itemize}
 \item[(iv)$_1$] Suppose that $a(t)$ is known as a~function of time.
 We~assume that this functional dependence is very slow (system $M$ is
 ``heavy''). It is then a~justified approximation to assume that the
 time-dependent hamiltonian of the evolution of the system $Q$ is
 given by $H_{a(t)}$ (with $H_a$ defined in (iii) above). As~$a(t)$ is
 slowly varying we assume the adiabatic approximation to calculate
 the evolution. One is usually interested in the situations in
 which the initial state of $Q$ is given by an eigenvector of~$H_a$
 for the initial value of~$a$. Suppose that for each $a$ we have
 a~nondegenerate, normalized eigenvector $\psi_a$ of $H_a$:
\begin{equation}\label{ext-eiga}
  H_a\psi_a=E_a\psi_a\,,
\end{equation}
 and the family $\psi_a$
 depends continuously on $a$. If at $t=0$ the state of $Q$ was given
 by $\psi_{a(0)}$, then at later times in the adiabatic approximation
 its state is equal to $\psi(t)= e^{i\varphi(t)} \psi_{a(t)}$,
 where $\varphi(t)$ is a~real function depending functionally on
 $E_a$ and $\psi_a$.
 If an operator $B$ represents an observable, then the time-dependence
 of its expectation value is given by
\begin{equation}\label{ext-mean}
 \<B\>_t=(\psi_{a(t)},B\,\psi_{a(t)})\,,
\end{equation}
 so it is a~function of $a$ in this approximation. It is important to
 note that the eigenvalues $E_a$ are modified by the addition of
 $\lambda_a$ under the transformation~(\ref{ext-gau}), but both the
 eigenvectors $\psi_a$ and the mean values $\<B\>_t$ remain
 unchanged.
 \item[(iv)$_2$] Finally, the evolution of the macroscopic variables
 $a(t)$ must be determined. This is the most controversial part of
 the problem, but we believe that the foregoing discussion indicates
 its proper solution.\\
 The intrinsic energy stored in the quantum part
 $Q$ is represented (in~the Schr\"odinger picture) by the operator
 $H$ ((i)$_3$ above), which in the coupled system is not a~constant
 of motion any more. Under the assumptions of (iv)$_1$ its
 expectation value is a~function of $a$, depending on the choice of
 the continuous family of eigenvectors $\psi_a$:
 \begin{equation}\label{ext-en}
  \mathcal{E}_a\df (\psi_a,H\,\psi_a)\,,
 \end{equation}
 and the time-dependence of this expectation value is through $a(t)$
 only. Changes in $\mathcal{E}_a$ correspond to the energy which has
 been transferred from $Q$ to the rest of the system, which (with the
 suppression of all microscopic details of $M$) is described by the
 variables $a$. Thus $\mathcal{E}_a$ plays the role of a~potential
 energy with respect to these variables. We~assume that the rest of
 the total energy of the coupled system is supplied by the kinetic
 energy of $M$, thus we obtain a~potential system, with the
 generalized force given by
\begin{equation}\label{ext-for}
  \mathcal{F}_a=-\frac{\partial \mathcal{E}_a}{\partial a}\,.
\end{equation}
 With a~specific form of the kinetic energy for a~particular model
 the motion of $a(t)$ could be determined, and with large
 inertial parameters (a ``heavy'' system)
 the approximation of its slow change should be confirmed.
 \end{itemize}
\end{itemize}

We~have thus spelled out all the assumptions and arrived at the
basic formulas~(\ref{ext-en}),~(\ref{ext-for}). In~the following
sections we shall take these formulas as a~starting point. The
derivation of the formulas was not rigorous, as this would demand
more information on the underlying microscopic model of the closed
system $Q\!-\!M$. A~detailed analysis of these questions is both
outside the usual discussions of Casimir effect, and also beyond
the reach of a~rigorous calculation at present. However, we
believe that the proposed discussion offers more plausibility than
most of the statements of the problem to be found in literature.
In~particular, points made by us in (ii) above are typically
ignored; we shall see their consequences when $Q$ is an
infinite-dimensional system, e.g.\ a~quantum field. Furthermore,
we want to draw a~closer attention to the formula~(\ref{ext-en})
and contrast it with what one obtains by the generalization of the
``zero point'' method to the more general context discussed in
this section. In~the latter case our formulas~(\ref{ext-en}) and
(\ref{ext-for}) are replaced respectively by
\begin{equation}\label{ext-focas}
  \mathcal{E}^{\mathrm{z.p.}}_a=E_a-E_0\,,\qquad
  \mathcal{F}^{\mathrm{z.p.}}_a
  =-\frac{\partial E_a}{\partial a}\,,
\end{equation}
where $E_a$ is the eigenvalue determined by~(\ref{ext-eiga}), and
$E_0$ some reference eigenvalue of $H$. One can object to these
formulas on several grounds.
 \begin{itemize}
 \item[(a)] The philosophy behind them seems to be this: the
 backreaction of $Q$ on $M$ is due to the changes in $H_a$, which may
 be interpreted as the sum of intrinsic energy $H$ of $Q$ and some
 interaction energy. However, we think that it is $M$ which absorbs
 the interaction and transforms it in a~phenomenological way into an
 effect on macroscopic variables $a$, while $Q$ has a~rather
 clear-cut identity.
 \item[(b)] The energy given by the ``zero point'' philosophy is not
 a~quantum mechanical average of any clear-cut observable: with
 changing $a$ one changes the observable $H_a$. Moreover, as already
 pointed out, $H_a$ and their eigenvalues are subject to the gauge
 freedom~(\ref{ext-gau}). The usual argument runs that this is fixed by
 the quantization of the ``proper'' classical expression for $H_a$.
 We~regard this argument as very unreliable. Quantum theory is the
 more fundamental one, so in case of doubt it should not seek
 a~verdict from the classical theory.
 \item[(c)] We~put forward the following
 ``consistency check''. Suppose that for certain values of parameters
 $a$ the effect of $M$ on $Q$ vanishes. In~this case the backreaction
 force should vanish as well. The supposition means that for $a=a_0$
 the vector $\psi_{a_0}$ is also an eigenvector of $H$,
 $H\psi_{a_0}=E\psi_{a_0}$ with some eigenvalue $E$.
 Using this equation one easily shows that our formulas yield
\begin{equation*}
 \mathcal{F}_{a_0}
 =-E\,\frac{\partial (\psi_a,\psi_a)}{\partial a}\bigg|_{a=a_0}
 =0\,,
\end{equation*}
 so they pass the check. On the other hand
 \begin{equation*}
 \mathcal{F}^{\mathrm{z.p.}}_{a_0}
  =-\frac{\partial E_a}{\partial a}\bigg|_{a=a_0}\,,
 \end{equation*}
 which, in general, has no reason to vanish.
 \item[(d)] In~Section~\ref{fd} below we discuss an example of
 a~Casimir-type effect in a~canonical system with finite degrees of
 freedom. In~this example the ``zero point'' method fails
 dramatically, yielding a~completely unphysical result.
 \end{itemize}
\noindent How, then, may ``zero point'' expressions arise?
We~shall show in Section~\ref{zem} below how for quantum fields
problems ``zero point'' expressions may be related to ours
 by unjustified idealizations and manipulations.

\setcounter{equation}{0}
\section{A~class of quasi-free systems}\label{cq}

We~discuss in this section a~general quantization scheme for
a~class of simple models. This class includes linear perturbations
of multi-dimensional harmonic oscillators or quantum fields.

Consider first the classical case. Let $\R$ be a~real Hilbert
space, and denote its scalar product by $(.\,,.)$. Let $h$ be
a~selfadjoint, strictly positive (hence invertible, with densely
defined inverse $h^{-1}$) operator in~$\R$, with the domain
$\D_\R(h)$. We~form the external direct sum
$\Lc=\D_\R(h)\oplus\R\subset\R\oplus\R$, and denote its elements
by $V=v\oplus u$, $v\in\D_\R(h)$, $u\in\R$. With the symplectic
form $\sigma$ defined by
\begin{equation}\label{cq-sym}
  \sigma(V_1,V_2)=(v_2,u_1)-(v_1,u_2)
\end{equation}
space $\Lc$ becomes the phase space of a~classical  model. Let the
Hamiltonian function of the model be given by
$\Hc(v,u)=\frac{1}{2}[(u,u)+(hv,hv)]$ (where all mass parameters
have been absorbed by momenta). The evolution determined by this
Hamiltonian in $\Lc$ is given by
\begin{equation}\label{cq-ev}
  T_t(v\oplus u)=\big(\cos(ht)v+\sin(ht)h^{-1}u\big)\oplus
  \big(-\sin(ht)hv+\cos(ht)u\big)\,.
\end{equation}
The differential form of this evolution is actually valid only on
a subspace of $\Lc$ (dense in $\R\oplus\R$), but the evolution
itself is properly defined on the whole of~$\Lc$. Operators $T_t$
form a~one-parameter group of symplectic transformations
\begin{equation}\label{cq-st}
  T_tT_s=T_{t+s}\,,\qquad \sigma(T_tV_1,T_tV_2)=\sigma(V_1,V_2)\,.
\end{equation}
Note, also, that
\begin{equation}\label{cq-tref}
  T_{-t}=(\id\oplus-\id)\,T_t\,(\id\oplus-\id)\,.
\end{equation}
Each $V'\in\Lc$ may be identified with an element of the dual
space by the rule
\begin{equation}\label{cq-fu}
  V'(V)=(v',u)+(u',v)=\sigma(V',(\id\oplus-\id)V)\,.
\end{equation}
Then using Eqs.\,(\ref{cq-st}) and~(\ref{cq-tref}) one easily
shows that
\begin{equation}\label{cq-clev}
  (T_tV')(V)=V'(T_tV)\,.
\end{equation}

The above model may be generalized by considering a~more general
subspace contained in $\D_\R(h)\oplus\R$ and invariant under the
evolution law~(\ref{cq-ev}). We~use this freedom to choose
\begin{equation}\label{cq-le}
  \Lc=\D_\R(h)\oplus\D_\R(h^{-1/2})
\end{equation}
(the invariance under~(\ref{cq-ev}) is easily checked). The
evolution law $T_t$ may be now expressed as a~unitary evolution in
a complex Hilbert space. One introduces a~complex Hilbert space
$\K$ which is the complexification of $\R$, $\K=\R\oplus\, i\R$,
with scalar product (denoted by the same symbol) and complex
conjugation defined by
\begin{gather}\label{cq-compl}
  (v_1+iu_1,v_2+iu_2)=(v_1,v_2)
  +(u_1,u_2)+i(v_1,u_2)-i(u_1,v_2)\,,\\
  \K\ni x\mapsto Kx\equiv\bar{x}\in\K\,,\quad K(v+iu)=v-iu\,.
\end{gather}
We~shall write $v=\Rp(v+iu)$, $u=\Ip(v+iu)$. The operator $h$ has
a unique extension to a~complex-linear operator on $\K$, denoted
by the same symbol, with the domain $\D(h)=\D_\R(h)\oplus
i\D_\R(h)$. This new $h$ is again a~selfadjoint, positive
operator, and it commutes with the conjugation. Consider now
a~real-linear operator
\begin{equation}\label{cq-jey}
j:\Lc\mapsto\Ran j\subset \K\,,\qquad j(V)=h^{1/2}v-ih^{-1/2}u\,.
\end{equation}
Its range $\Ran j$ is a~real-linear subspace of $\K$, dense in
$\K$, and $j$ is a~bijection of $\Lc$ onto $\Ran j$. Then for all
$V\in\Lc$:
\begin{equation}\label{cq-harm}
 j(T_tV)=e^{iht}j(V)\,,
\end{equation}
so $\Ran j$ is invariant under $e^{iht}$, and the evolution may be
expressed as
\begin{equation}\label{cq-evh}
  T_tV=j^{-1}(e^{iht}j(V))\,.
\end{equation}

Space $\K$, regarded as a~real vector space, has a~natural
symplectic structure introduced with the symplectic form
$\Ip\,(f,g)$. Space $\Ran j$ is its symplectic subspace. One
easily shows that
\begin{equation}\label{cq-srj}
  \sigma(V_1,V_2)=\Ip\,(j(V_1),j(V_2))\,,
\end{equation}
so $j$ is a~symplectic transformation of $\Lc$ onto $\Ran j$.
\pagebreak

The mapping $j$, as well known, serves to construct the ground
state representation of the quantum version of the model, and the
space $\K$ is then the ``one-particle space'' (see below).
A~natural problem thus arises: to extend the construction of the
space $\Lc$ to the largest possible space compatible with the
symplectic mapping~(\ref{cq-jey}), that is to extend $\Lc$ and $j$
so as for
 $\Ran j$ to cover the whole space $\K$ (instead of being only dense in
$\K$, as above). One defines on $\D_\R(h^{\pm 1/2})$ the scalar
products
\begin{align}
  (v_1,v_2)_+&=(h^{1/2}v_1,h^{1/2}v_2)\,,&\quad
  &v_1,v_2\in\D_\R(h^{1/2})\,,\label{cq-sppl}\\
  (u_1,u_2)_-&=(h^{-1/2}u_1,h^{-1/2}u_2)\,,&\quad
  &u_1,u_2\in\D_\R(h^{-1/2})\,,\label{cq-spmi}
\end{align}
and denotes by $\R_+$ and $\R_-$ the Hilbert spaces obtained by
the completion of $\D_\R(h^{1/2})$ and $\D_\R(h^{-1/2})$,
respectively, with respect to the norms $\|v\|_+=\sqrt{(v,v)_+}$
and $\|u\|_-=\sqrt{(u,u)_-}$. For $v\in\D_\R(h^{1/2})$ and
$u\in\D_\R(h^{-1/2})$ we have \linebreak $\|h^{1/2}v\|=\|v\|_+$
and $\|h^{1/2}u\|=\|u\|_-$. Therefore operators $h^{1/2}$ and
$h^{-1/2}$ extend by continuity to bijective isometric operators
$\wh{h^{1/2}}$ and $\wh{h^{-1/2}}$ respectively,
\begin{gather}\label{cq-ext}
    \wh{h^{1/2}}:\R_+\mapsto\R\,,\qquad
    \|\wh{h^{1/2}}v\|=\|v\|_+\,,\\
    \wh{h^{-1/2}}:\R_-\mapsto\R\,,\qquad
    \|\wh{h^{-1/2}}u\|=\|u\|_-\,.
\end{gather}

We~note for future use that
\begin{equation}\label{cq-com}
    \R_\pm\cap\R=\D_\R(h^{\pm1/2})\,.
\end{equation}
This is easily seen in the spectral representation of $h$: if $h$
is a~multiplication by a~positive, different from zero almost
everywhere, function $f$ in a~space $L^2(M,d\mu)$, then $\R$
consists of functions $\psi$ for which
$\int_M|\psi(m)|^2d\mu(m)<\infty$, $\R_\pm$ consists of functions
for which $\int_M(f(m))^{\pm1}|\psi(m)|^2d\mu(m)<\infty$, and
$\D_\R(h^{\pm1/2})$ -- of those satisfying both conditions.

For $v\in\D_\R(h^{1/2})$ and $u\in\D_\R(h^{-1/2})$ one has
\begin{equation}\label{cq-extpair}
    |(v,u)|=|(h^{1/2}v,h^{-1/2}u)|\leq\|v\|_+\|u\|_-\,,
\end{equation}
thus $(v,u)$ extends to a~continuous pairing
\begin{equation}\label{cq-pairing}
    \R_+\times\R_-\ni v,u\mapsto \<v,u\>\in\mR\,,\quad
    |\<v,u\>|\leq\|v\|_+\|u\|_-\,.
\end{equation}
Now one can set
\begin{gather}
    \wh{\Lc}=\R_+\oplus\R_-\,,\quad
    \wh{\sigma}(V_1,V_2)=\<v_2,u_1\>-\<v_1,u_2\>\,,\label{cq-largest}\\
    \wh{j}:\wh{\Lc}\mapsto\K\,,\quad
    \wh{j}(V)=\wh{h^{1/2}}v-i\wh{h^{-1/2}}u\,,\label{cq-jlargest}\\
    \wh{T}_tV=\wh{j}^{-1}(e^{iht}\wh{j}(V))\,.\label{cq-tlargest}
\end{gather}
As~a~consequence of~(\ref{cq-com}) one has
\begin{equation}\label{cq-comlr}
    \wh{\Lc}\cap(\R\oplus\R)=\D_\R(h^{1/2})\oplus\D_\R(h^{-1/2})\,.
\end{equation}
It is easy to see that now $\Ran\wh{j}=\K$, the space given by
Eq.\,(\ref{cq-le}) is dense in $\wh{\Lc}$ (in its Hilbert space
structure norm), and the time evolution on $\wh{\Lc}$ is the
continuous extension of the evolution on the space~(\ref{cq-le}).
Moreover,
\begin{equation}\label{cq-jscal}
    (\wh{j}(V_1),\wh{j}(V_2))=
    (v_1,v_2)_++(u_1,u_2)_-+i\wh{\sigma}(V_1,V_2)\,,
\end{equation}
so, in particular, $\wh{j}$ is a~symplectic mapping of
$(\wh{\Lc},\wh{\sigma})$ onto $(\K,\Ip(.,.))$. Relations
(\ref{cq-fu}) and~(\ref{cq-clev}) are also generalized to
\begin{gather}
  V'(V)=\<v',u\>+\<u',v\>=\wh{\sigma}(V',
  [\id\oplus(-\id)]V)\,,
  \label{cq-fugen}\\
  (\wh{T}_tV')(V)=V'(\wh{T}_tV)\,.\label{cq-clevgen}
\end{gather}
Once we have the largest arena consistent with the scheme,
particular models are defined by choosing a~subspace invariant
under the evolution:
\begin{equation}\label{cq-small}
    \Lc\subset\wh{\Lc}\,,\qquad \wh{T}_t\Lc\subset\Lc\,.
\end{equation}
The maximal model is invariant under the time reversal,
represented by the operator $\id\oplus(-\id)$ appearing in
(\ref{cq-fugen}). We~want to retain this property for the model
defined by $\Lc$, which is equivalent to the assumption
\begin{equation}\label{cq-tr}
    \Lc=\Lc_+\oplus\Lc_-\,,\quad \Lc_\pm\subset\R_\pm\,.
\end{equation}

Examples of particular spaces include the class of spaces
\begin{equation}\label{cq-lpart}
  \begin{split}
  \Lc&=\D_\R(h^{r+1})\cap\D_\R(h^{-s})\oplus
  \D_\R(h^r)\cap\D_\R(h^{-t-\frac{1}{2}})\,,\\
  &\hspace{13pt}r,s,t\in\<0,\infty)\,,\quad s\leq t+\tfrac{1}{2}\,,
  \quad t\leq s+\tfrac{3}{2}\,,
  \end{split}
\end{equation}
all of which are contained in~(\ref{cq-le}).

The quantum version of the maximal model (with the symplectic
space $\wh{\Lc}$) is now obtained by standard procedure (see e.g.
\cite{br}, vol.\,II). Starting with expression~(\ref{cq-fugen})
one aims at replacing $V'$ by some ``quantum variable'' $\Phi$.
In~quantum theory a~concrete representation of a~quantum variable
is an operator in a~Hilbert space. If the classical variable is
real, its quantum counterpart should be represented by
a~selfadjoint operator. Thus one assumes that a~Hilbert space
$\Hc$ is given, and for each $V\in\Lc$ one has a~selfadjoint
operator $\Phi(V)$ in that space. The functional dependence of
$\Phi(V)$ on $V$ is assumed to be linear, and the canonical
commutation relations are imposed:
\begin{equation}\label{cq-ccr}
  [\Phi(V_1),\Phi(V_2)]=i\wh{\sigma}(V_1,V_2)\id\,,
\end{equation}
where one still has to clarify the domain problems. If $V=v\oplus
u$, then we shall also write $\Phi(V)=\Phi(v,u)$. The element
$P(v)\equiv\Phi(v,0)$ has the interpretation of the quantum
momentum for the ``test vector''~$v$, and $X(u)\equiv\Phi(0,u)$ --
of the quantum position variable for the ``test vector''~$u$. With
the linearity of $\Phi(V)$ the above commutation relations are
equivalent to those in a~more familiar form
\begin{equation}\label{cq-ccrf}
  [X(u_1),X(u_2)]=0\,,\ [P(v_1),P(v_2)]=0\,,\
  [P(v),X(u)]=-i\<v,u\>\id\,.
\end{equation}

It is well-known that there are many different concrete
representations of the above scheme, and this is why it is
desirable to formulate the canonical commutation relations in an
algebraic way. As~there are no bounded operators satisfying these
relations, it is usual to take them in an exponentiated variant.
This leads to the Weyl form of these relations. The Weyl algebra
over the symplectic space $\wh{\Lc}$ is the unique $C^*$-algebra
generated by elements $W(V)$, $V\in\wh{\Lc}$, and a~unit
element~$\1$, by the relations
\begin{equation}\label{cq-weyl}
 \begin{split}
  W(V_1)&W(V_2)=e^{\textstyle -\tfrac{i}{2}\wh{\sigma}(V_1,V_2)}
  \,W(V_1+V_2)\,,\\
  &W(V)^*=W(-V)\,,\qquad W(0)=\1\,.
 \end{split}
\end{equation}
One looks for representations of this algebra by bounded operators
in a~Hilbert space (which exist for all $C^*$-algebras). Let $\pi$
be such a~representation in the Hilbert space $\Hc$. One says that
this representation is regular, if for each $V\in\wh{\Lc}$ the
one-parameter group of unitary operators
\begin{equation}\label{cq-reg}
  \mathbb{R}\ni s\mapsto \pi(W(sV))
\end{equation}
is strongly continuous. If this is the case, then there exist, by
Stone's theorem (e.g.~\cite{rs}), selfadjoint operators $\Phi(V)$
such that
\begin{equation}\label{cq-exp}
 \pi(W(V))=\exp(i\Phi(V))\,.
\end{equation}
Moreover, one shows that for each finite-dimensional subspace
$\Lc'\subset\wh{\Lc}$ there exists a~dense subspace
$\D'\subset\Hc$ which is contained in the domains of all operators
$\Phi(V)$, $V\in\Lc'$, is an invariant subspace and an essential
domain of selfadjointness for all of them, and on which linearity
of $\Phi(V)$ in its argument $V\in\Lc'$ and commutation relations
(\ref{cq-ccr}) are satisfied (this follows from the Stone--von
Neumann uniqueness theorem, cf.~\cite{br}, vol.II). While not all
canonical systems with these properties arise in this way from
regular representations of the corresponding Weyl algebra, most of
those needed in physics do, and one usually restricts attention to
this class.

The algebra $\A$ of the maximal model is thus the Weyl algebra
over $\wh{\Lc}$. Dynamics of the model is a~``quasi-free''
evolution obtained by a~simple ``quantization'' of the classical
evolution $\wh{T}_t$. Being guided by the replacement \mbox{$V'\to
\Phi$} in Eq.\,(\ref{cq-clevgen}) and the relation~(\ref{cq-exp}),
one defines it on the algebraic level by
\begin{equation}\label{cq-qev}
  \alpha_t(W(V))=W(\wh{T}_tV)\,.
\end{equation}
One looks now for a~representations $\pi$ of the algebra in which
this evolution law may be implemented:
\begin{equation}\label{cq-uev}
  \pi(W(\wh{T}_tV))=U(t)\,\pi(W(V))\,U(t)^*\,,\quad
  U(t)=e^{\textstyle itH}\,,
\end{equation}
where $H$ is a~selfadjoint operator. The ground state
representation is obtained if $H$ is a~nonnegative operator with
zero energy ground state. This representation is constructed in
standard way with the use of the Fock space method. \linebreak Let
$\WF(f)=\exp[i\FF(f)]$, $f\in\K$, be the Weyl system of operators
in the Fock space~$\Hc$ built on the ``one-particle'' space $\K$
(see Appendix~\ref{we}). We~set
\begin{equation}\label{cq-rep}
  \pi(W(V))\df\WF(\wh{j}(V))\,,
\end{equation}
or, which is equivalent,
\begin{equation}\label{cq-repf}
  \pi(W(V))=e^{\textstyle i\Phi(V)}\,,\qquad
  \Phi(V)=\FF(\wh{j}(V))\,.
\end{equation}
Using identity~(\ref{cq-jscal}) and properties of the operators
$\WF(f)$ one easily shows that this indeed constitutes
a~representation of the Weyl algebra~(\ref{cq-weyl}). By the
irreducibility of the Weyl system in Fock space this
representation is irreducible. Moreover, using
Eqs.\,(\ref{cq-rep}) and~(\ref{cq-tlargest}) one rewrites the
condition~(\ref{cq-uev}) as
\begin{equation}
 \WF(e^{ith}f)=e^{\txt itH}\WF(f)e^{\txt -itH}\,,\quad f\in\Hc\,.
\end{equation}
The discussion of Appendix~\ref{we} shows now that
\begin{equation}\label{cq-ham}
  H=d\Gamma(h)\,,
\end{equation}
where $d\Gamma(h)$ is the ``second quantization'' of $h$ (see
Eqs.\,(\ref{we-dgh} --~\ref{we-esdg})). This energy operator has
nonnegative spectrum, and a~unique ground state represented by the
``Fock vacuum'' $\Omega$.

Consider now a~restriction of this model defined by a~subspace
$\Lc$ invariant under evolution (Eq.\,(\ref{cq-small})) and time
reflection (Eq.\,(\ref{cq-tr})). The algebra of the model is the
subalgebra of the Weyl algebra~(\ref{cq-weyl}) obtained by
restricting the test vectors to $\Lc$. It is well-known, that the
resulting model is not identical with the maximal one if
$\Lc\neq\wh{\Lc}$ (see~\cite{br}, vol.\,II), but one can demand
that its ground state representation approximates that of the
maximal model. This representation may be constructed as before,
but the scope of Weyl operators used in this representation is
restricted to $\{\WF(f)\mid f\in\wh{j}(\Lc)\}$ -- cf.\
Eq.\,(\ref{cq-rep}). This set is irreducible in $\K$ if, and only
if, the space $\wh{j}(\Lc)$ is dense in $\K$, or, what is the
same, $\Lc$ is dense in $\R_+\oplus\R_-$. With the time reflection
symmetry assumption (\ref{cq-tr}) this takes the form
\begin{equation}\label{cq-dense}
    \Lc_+\ \text{is dense in}\ \R_+\,,\quad
    \Lc_-\ \text{is dense in}\ \R_-\,.
\end{equation}
We~restrict attention to those spaces $\Lc$ which satisfy this
condition. This restriction can be paraphrased by saying that
there are no superselection rules in the Fock space of the ground
state representation.
 \vspace{4ex}

 Examples
 \begin{itemize}
 \item[(i)] Multidimensional harmonic oscillator\\
 In~this case $\R$ is a~finite-dimensional Euclidean space, and $h$
 is a~positive selfadjoint operator defined on the whole of $\R$.
 We~choose $\Lc=\R\oplus\R$. Space~$\K$ is the unitary space obtained by
 complexification of $\R$, and $\Ran j=\K$. The more familiar simple
 form of the model is obtained by choosing in $\R$ an orthonormal basis
 $(e_1,\ldots,e_n)$ of eigenvectors of $h$ and putting
 \mbox{$X_i=\Phi(0,e_i)$}, $P_i=\Phi(e_i,0)$. The system is then the set of
 $n$ independent harmonic oscillators with canonical variables
 $\{X_i,P_i\}$, unit masses, and frequencies $\w_i$, where
 $he_i=\w_ie_i$.
 \item[(ii)] Free scalar field\\
 Free quantum fields are usually defined as operator-valued
 distributions on test functions of all spacetime variables. The
 evolution equation (Klein-Gordon for the scalar field) is already
 encoded in this formulation, which is manifestly relativistically
 covariant. For our purposes the equivalent initial value formulation
 is preferable -- we want to separate evolution law, as~far as it is
 possible, from setting up of the algebra.

 Standard identifications for this model are as follows:
 \begin{gather*}
 \R=L^2_\mR(\mR^3)\,,\ \ \K=L^2(\mR^3)\,,\label{cq-frk}\\
 h=\sqrt{-\Delta}\,,\label{cq-fh}\\
 \Lc=\D_\mR(\mR^3)\oplus\D_\mR(\mR^3)\,,\label{cq-fl}
 \end{gather*}
 where subscript $\mR$ denotes the real part of the respective
 function space, $\D(\mR^3)$ is the space of infinitely
 differentiable complex functions of compact support, and $\Delta$ is the
 Laplace operator. Standard solution of the initial value problem for
 the Klein-Gordon equation has the form of Eq.\,(\ref{cq-ev}), and
 the assumptions~(\ref{cq-small}),~(\ref{cq-tr}) and~(\ref{cq-dense}) are satisfied.
 \item[(iii)] Scalar field with external time-independent interaction\\
 Loosely speaking, the choice of $h$ here is the square root of
 a~selfadjoint positive operator of heuristic form
 ``$h^2=-\Delta+\text{interaction}$''. There are a~few
 possibilities.
 \begin{itemize}
 \item[(iii)$_1$] If the interaction is given by an external field
 $\sigma=\sigma(\vec{x})$ then the choice of spaces $\R$ and $\K$ remains
 the same as in the free case, while $h^2=-\Delta+\sigma$ \linebreak (we assume
 that $h^2$ is still positive -- there are no bound states).\linebreak Depending on
 the form of $\sigma$ the choice of $\Lc$ as in the free case may be
 admissible (satisfy assumptions~(\ref{cq-small}) and~(\ref{cq-dense})) or
 not. A~safe choice for $\Lc$ is supplied by any of the cases
 given by Eq.\,(\ref{cq-lpart}). More
 generally, $h^2$ may be any positive selfadjoint perturbation of $-\Delta$ in
 the sense of operators or forms.
 \item[(iii)$_2$] The next possibility arises from restricting the region
 accessible to the field to a~proper subset $\Lambda\subset\mR^3$.
 In~this case $\R=L^2_\mR(\Lambda)$, $\K=L^2(\Lambda)$, and
 $h^2=-\Delta_B$, where $\Delta_B$ is a~selfadjoint extension of the
 Laplace operator defined on twice differentiable functions with
 support inside~$\Lambda$, determined by some boundary conditions
 ``$B$''. Here, of course, the free field choice of $\Lc$ is not
 admissible, and a~safe choice is again given by the formula
~(\ref{cq-lpart}).
 \item[(iii)$_3$] Finally, we consider a~setting usually assumed for
 the Casimir effect. The whole physical space $\mR^3$ is divided by
 two-dimensional surfaces into disjoint open regions
 $\Lambda_1,\ldots,\Lambda_s$. Position of the dividing boundaries
 is characterized by a~set of parameters $a$. One chooses $\R$ and
 $\K$ as in the free field case. Depending on parameters $a$,
 a~family of positive operators $h_a$ is given by $h^2_a=-\Delta_a$,
 where $\Delta_a$ is the Laplace operator in $L^2(\mR^3)$ determined
 by the assumed boundary conditions (Dirichlet, Neumann, etc.) at the
 dividing surfaces with positions given by parameters $a$.
 In~consequence, the choice of the symplectic space $\Lc$ must be
 adjusted to the position of the boundaries. A~choice of simple
 possibilities is
 again given by Eq.\,(\ref{cq-lpart}) (with $h_a$ replacing $h$). In~the
 Casimir problem one wants to compare states of the system at
 different values of $a$. However, spaces $\Lc$ depend nontrivially
 on $a$, thus the respective Weyl algebras are also different, and do
 not define the same quantum system. This constitutes the difficulty
 of traditional treatments of the Casimir effect which we anticipated
 in Section~\ref{ext}. Infinities naturally appear then, and are
 a~consequence of an uncritical use of the notion of a~quantum field.
 \end{itemize}
 \end{itemize}
In~the following sections we discuss Casimir effects for some
systems in the category described in the present section. We~start
with a~simple finite-di\-men\-sio\-nal case.

\setcounter{equation}{0}
\section{Deformation of a~finite-dimensional harmonic oscillator}\label{fd}

Our unperturbed quantum system $Q$ is here an $n$-dimensional
quantum oscillator described in example (i) of the last section.
Thus the algebra of the model is the Weyl algebra based on
a~finite-dimensional symplectic space $\Lc=\R\oplus\R$, and its
representation is given by $\pi(W(V))=\WF(j(V))$ in the Fock space
$\Hc$ based on the finite-dimensional one-excitation space $\K$.
The energy operator of the model is given by $H=d\Gamma(h)$.

We~consider now a~combined system $Q\!-\!M$, as described in
Section~\ref{ext}, and assume that the influence of $M$ on $Q$ for
frozen parameters $a$ manifests itself in the change of axes and
frequencies of oscillations. Thus the time evolution of the
algebra for frozen parameters is given by
\begin{equation}\label{fd-eva}
  \alpha_{at}(W(V))=W(T_{at})\,,
\end{equation}
where $T_{at}$ has the form~(\ref{cq-ev}), but with operator $h$
replaced by an operator $h_a$ from a~family $\{h_a\}$. For each
$a$ the irreducible representation of the algebra in which this
evolution is implemented by a~unitary one-parameter group, with
a~nonnegative energy operator, is constructed by the same method,
as in the free $Q$ case, in the same Fock space $\Hc$. Thus
\begin{equation}\label{fd-def}
  j_a(V)=h_a^{1/2}v-ih_a^{-1/2}u\,,\qquad \pi_a(W(V))=\WF(j_a(V))\,.
\end{equation}
The Hamilton operator is given by $d\Gamma(h_a)$, with ground
state described by the Fock vacuum $\Omega$. However, as discussed
in Section~\ref{ext}, we want to describe the same physical
situation with the use of the representation $\pi$. Therefore for
each $a$ we look for a~unitary operator $U_a$ which by similarity
transforms representation $\pi_a$ onto $\pi$:
\begin{equation}\label{fd-uer}
  U_a\pi_a(W(V))U_a^*=\pi(W(V))\,,\quad
  U_a\Phi_a(V)U_a^*=\Phi(V),\qquad V\in\Lc\,.
\end{equation}
We~substitute here the definitions of the representations $\pi$
and $\pi_a$, and set \linebreak $j_a(V)\equiv f$. This condition
then takes the form
\begin{equation}\label{fd-uef}
 U_a\WF(f)U_a^*=\WF(L_af)\,,\quad f\in\K\,,
 \quad\text{where}\ \ L_a\df j\,j_a^{-1}\,.
\end{equation}
Both $j$ and $j_a$ are bijective symplectic mappings, so $L_a$ is
a~symplectic transformation of the space $(\K,\Ip(.\,,.))$, and
the above condition states that $U_a$ implements the corresponding
Bogoliubov transformation in the Fock space (see
Appendix~\ref{bo}). As~$\K$ is finite-dimensional, such $U_a$
exists. The explicit form of transformation $L_a$ is easily
obtained:
\begin{equation}\label{fd-L}
  L_af=h^{1/2}h_a^{-1/2}\Rp f+ih^{-1/2}h_a^{1/2}\Ip f\,,
\end{equation}
and then from~(\ref{st-ts}) one finds $L_a=T_a+S_a$, $T_a$
complex-linear and $S_a$ complex-antilinear,
\begin{equation}\label{fd-ts}
 T_a=\frac{1}{2}(B_a^{-1}+B_a^*)\,,\quad
 S_a=\frac{1}{2}(B_a^{-1}-B_a^*)K \,.
\end{equation}
where
\begin{equation}\label{fd-kc}
  B_a=h_a^{1/2}h^{-1/2}\,,\qquad Kf=\bar{f}\,.
\end{equation}

In~the representation $\pi$ the Hamiltonians of the new evolutions
are given by
\begin{equation}\label{fd-nham}
  H_a=U_ad\Gamma(h_a)U_a^*\,,
\end{equation}
and the ground state of $H_a$ is given by
\begin{equation}\label{fd-gr}
  \Omega_a=U_a\Omega\,.
\end{equation}

Suppose now, that $a$ is a~single real parameter, and under the
influence of the external conditions the state of the subsystem
$Q$ changes adiabatically over the states $\Omega_a$, as discussed
in Section~\ref{ext}. The potential for the backreaction force is
therefore, in accordance with point (iv) in Section~\ref{ext},
determined by
\begin{equation}\label{fd-ce}
  \E_a=(\Omega_a,H\,\Omega_a)\,.
\end{equation}
We~take into account that $H=d\Gamma(h)$ and use the expression
for a~form matrix element of $d\Gamma(h)$ as given in
Eqs.\,(\ref{we-qf}),~(\ref{we-qfo}):
\begin{equation*}
 \E_a=\sum_i\|a(h^{1/2}f_i)\Omega_a\|^2\,,
\end{equation*}
where $\{f_i\}$ is an arbitrary orthonormal basis of $\K$. We~use
Eqs.\,(\ref{bo-abo}),~(\ref{bo-acom}) and~(\ref{bo-an}) to find
\begin{equation*}
 \|a(h^{1/2}f_i)\Omega_a\|^2=\|a_{L_a}^*(S_a{}^*h^{1/2}f_i)\Omega_a\|^2=
 (f_i,h^{1/2}S_aS_a{}^*h^{1/2}f_i)\,.
\end{equation*}
Thus we obtain
\begin{equation}\label{fd-res}
 \E_a=\Tr\big[h^{1/2}S_aS_a{}^*h^{1/2}\big]=
 \frac{1}{4}\Tr\big[(h_a-h)h_a^{-1}(h_a-h)\big]\,.
\end{equation}

Let the eigenvalues and orthonormal eigenvectors of $h$ and $h_a$
be given by
\begin{equation}\label{fd-eig}
 he_i=\epsilon_i e_i\,,\quad h_ae_{ai}=\epsilon_{ai}e_{ai}\,.
\end{equation}
Then using the spectral representation
 $h_a^{-1}=\sum_k\epsilon_{ak}^{-1}|e_{ak}\rangle\langle e_{ak}|$ and employing
the basis $e_i$ for the calculation of the trace we find
\begin{equation}\label{fd-ced}
 \E_a=\sum_{i,k}
 \frac{(\epsilon_{ak}-\epsilon_i)^2}{4\,\epsilon_{ak}}
 |(e_{ak},e_i)|^2\,.
\end{equation}

We~consider a~simple example. Let $Q$ be a~two-dimensional
oscillator in physical space, vectors $e_1,e_2$ being its main
axes, and let the effect of the external conditions be the
rotation of these axes by an angle $\varphi$ ($\equiv a$), without
a change in the frequencies. In~this case we have
$\epsilon_{\varphi k}=\epsilon_k$,
 $(e_{\varphi 1},e_2)=-(e_{\varphi 2},e_1)=\sin\varphi$, so
\begin{equation}
 \E_\varphi=\frac{(\epsilon_2-\epsilon_1)^2}{4}
 (\epsilon_1^{-1}+\epsilon_2^{-1})\sin^2\varphi\,.
\end{equation}
The backreaction ``force'' in this case is a~torque
\begin{equation}
 \F_\varphi=-\frac{(\epsilon_2-\epsilon_1)^2}{4}
 (\epsilon_1^{-1}+\epsilon_2^{-1})\sin 2\varphi\,.
\end{equation}
For $\epsilon_1\to 0$ (with $\epsilon_2$ kept constant) the torque
tends to infinity. This is what one should expect. This limiting
case describes the situation in which the harmonic force in the
direction of $e_2$ extends translationally invariant in the
direction of $e_1$; any rotation of this picture involves an
``infinite'' change.

Note, that the ``zero point'' prescription for the force gives
zero in the above example, which is an utterly unphysical
prediction.

\setcounter{equation}{0}
\section{An infinite-dimensional system}\label{id}

Let now $\R$ be an infinite dimensional real Hilbert space.
We~want to consider a~situation analogous to that discussed in the
last section: system $Q$ defined by the operator $h$, and its
perturbations by a~family of operators $h_a$. We~need to take into
account complications arising from the unboundedness of the
operators, as~explained in Section~\ref{cq}.

We~want to be able to define for our model both evolutions (that
determined by $h$ and by $h_a$), and both ground state
representations. Thus the model has to fit into structures defined
in Section~\ref{cq} both by $h$ as well as $h_a$. In~particular,
its symplectic space should in a~canonical way be a~part of both
$\wh{\Lc}$ and $\wh{\Lc}_a$. However, the construction of these
spaces is based on different, in general, parts of $\R\oplus\R$
($\D_\R(h^{1/2})\oplus\D_\R(h^{-1/2})$ and
$\D_\R(h_a^{1/2})\oplus\D_\R(h_a^{-1/2})$ respectively), and
without some restrictions there is no canonical way of
identification of their parts. We~assume that
\begin{equation}\label{id-dense}
    \D_\pm\equiv\D_\R(h^{\pm1/2})\cap\D_\R(h_a^{\pm1/2})\quad
    \text{is dense in}\quad \R_\pm\ \text{and in}\ \R_{a\pm}\,,
\end{equation}
(in this, and similar statements below, signs are either all
upper, or all lower).    In~this case space $\R_\pm$ is the
completion of $\D_\pm$ with respect to the norm $\|.\|_\pm$, and
$\R_{a\pm}$ is the completion of the same subspace with respect to
$\|.\|_{a\pm}$.

Suppose now that the space of a~model contains at least a~subspace
$\Lc^0$ such that
\begin{equation}
   \Lc^0=\Lc^0_+\oplus\Lc^0_-\,,\quad \Lc^0_\pm\subset\D_\pm\,,\quad
   \Lc^0_\pm\ \text{is dense in}\ \R_\pm\ \text{and in}\
   \R_{a\pm}\,,
\end{equation}
which is a~strengthening of the condition~(\ref{cq-dense}). Note
that then $\wh{j}(\Lc^0)=j(\Lc^0)$, $\wh{j}_a(\Lc^0)=j_a(\Lc^0)$,
and both spaces are dense in $\K$. Under these assumptions we show
that the following three conditions are equivalent:
\begin{itemize}
 \item[(i)] The symplectic mapping
\begin{equation}\label{id-sympl}
    L_a\df jj_a^{-1}:j_a(\Lc^0)\mapsto j(\Lc^0)
\end{equation}
extends to a~bounded operator in $\K$, with a~bounded inverse.
 \item[(ii)] The operators $h$ and $h_a$ satisfy the conditions
\begin{gather}
    \D_\R(h^{\pm1/2})=\D_\R(h_a^{\pm1/2})\,,\label{id-hha}\\
    B_a\equiv h_a^{1/2}h^{-1/2}\ \text{and}\ B_a^{-1}\
    \text{extend to bounded operators in}\ \K\,.
    \label{id-equiv}
\end{gather}
\item[(iii)] There exists a~selfadjoint, positive, bounded
operator $C_a$ in $\R$, with bounded inverse, and such that
\begin{equation}\label{id-haform}
    h_a=h^{1/2}C_ah^{1/2}
\end{equation}
in the sense of forms, that is: $h_a$ is the unique selfadjoint
operator defined by the closed form
$q(v_1,v_2)=(h^{1/2}v_1,C_ah^{1/2}v_2)$ with the form domain
$Q(q)=\D_\R(h^{1/2})$.
\end{itemize}
If these conditions are satisfied, then $\wh{\Lc}_a=\wh{\Lc}$, so
both evolutions are well defined in $\wh{\Lc}$.

Note that (i) is a~necessary condition for the ground state
representations defined by $h$ and $h_a$ to be equivalent. The
equivalence implies that Eq.\,(\ref{fd-uer}) is satisfied in
particular for all $V\in\Lc^0$, or, equivalently,
Eq.\,(\ref{fd-uef}) for all $f\in j_a(\Lc^0)$. But~then, as shown
in the Appendix~\ref{bo}, $L_a$ extends to a~bounded symplectic
mapping in $\K$, with a~bounded inverse.

Let (i) be satisfied. Then on $\Lc^0$ from Eq.\,(\ref{id-sympl})
we have $L_aj_a=j$ and \linebreak $j_a=L_a^{-1}j$, which implies
that for $w_\pm\in\Lc^0_\pm$ one has
\begin{equation}\label{id-eq}
    \|h^{\pm1/2}w_\pm\|\leq \con\,\|h_a^{\pm1/2}w_\pm\|\,,\quad
    \|h_a^{\pm1/2}w_\pm\|\leq \con\,\|h^{\pm1/2}w_\pm\|\,.
\end{equation}
This means that the norms $\|.\|_\pm$ and $\|.\|_{a\pm}$ are
equivalent on $\Lc^0_\pm$, so they yield the same completion, on
which these inequalities are preserved. But~$\Lc^0_\pm$ is dense
in $\R_\pm$ and $\R_{a\pm}$, so $\R_\pm=\R_{a\pm}$ as sets, hence
also $\wh{\Lc}=\wh{\Lc}_a$ as sets, with equivalent norms.
Moreover, from Eq.\,(\ref{cq-com}) we find
\[
 \D_\R(h_a^{\pm1/2})=\R_{a\pm}\cap\R=\R_\pm\cap\R=\D_\R(h^{\pm1/2})\,.
\]
The boundedness of $B_a$ and $B_a^{-1}$ follows now from
(\ref{id-eq}), which ends the proof of~(ii). Conversely, if (ii)
is satisfied, then one easily shows that the formula\linebreak
$L_af=B_a^{-1}\Rp f+iB_a^*\Ip f$ gives the extension needed in (i)
(cf.\ Eq.\,(\ref{fd-L})).

The equivalence of (ii) and (iii) follows by polar decomposition
of closed operators (e.g.~\cite{rs}). If we assume (ii), then
$h_a^{1/2}=B_ah^{1/2}$ with the domain $\D_\R(h^{1/2})$ is
a~selfadjoint operator, so $h_a=h^{1/2}B_a^*B_ah^{1/2}$, and
$C_a=B_a^*B_a$ fulfills the conditions of (iii). Conversely, let
$C_a$ be bounded, positive selfadjoint, with a~bounded inverse.
Then $C^{1/2}_ah^{1/2}$ with the domain equal to $\D_\R(h^{1/2})$
is a~closed operator. Indeed, let $v_n\in\D_\R(h^{1/2})$,
$\|v_n-v_m\|\to0$ and $\|C_a^{1/2}h^{1/2}(v_n-v_m)\|\to0$.
But~$C_a^{-1/2}$ is bounded, so also $\|h^{1/2}(v_n-v_m)\|\to0$.
As~$h^{1/2}$ is closed, there exists $v\in\D_\R(h^{1/2})$ such
that $\|v_n-v\|\to0$, $\|h^{1/2}(v_n-v)\|\to0$, and by boundedness
of $C_a^{1/2}$ also $\|C_a^{1/2}h^{1/2}(v_n-v)\|\to0$, which shows
that $C_a^{1/2}h^{1/2}$ is indeed closed. Thus the form $q$
defined in (iii) is closed, and the condition~(\ref{id-haform})
means that $|C_a^{1/2}h^{1/2}|=h_a^{1/2}$. It follows that there
exists an orthogonal operator $F_a$ such that
$C_a^{1/2}h^{1/2}=F_ah_a^{1/2}$, so $h_a^{1/2}h^{-1/2}$ extends to
a bounded operator $B_a=F_a^*C_a^{1/2}$. As~$C_a$ has a~bounded
inverse, so the same is true for $B_a^{-1}$, which ends the proof
of equivalence of (i)~--~(iii).

These preliminary results show that if the two ground state
representations are to be equivalent in our model, we have to
assume that~(\ref{id-haform}), and then all conditions (i) --
(iii), are true. Then the space $\wh{\Lc}$ is invariant under both
evolutions and forms the widest possible space of the model. Any
subspace (if it exists) $\Lc=\Lc_+\oplus\Lc_-\subset\wh{\Lc}$
which is also invariant under both evolutions and dense in
$\wh{\Lc}$ can also be taken as the symplectic space of the model.

With these assumptions the symplectic mapping $L_a$ decomposes
into the bounded complex-linear and complex-antilinear parts,
$L_a=T_a+S_a$. The application of the results described in
Appendix~\ref{bo} shows that the necessary and sufficient
condition for the unitary equivalence of the ground state
representations is that $S_a$ is a~Hilbert-Schmidt (HS) operator,
that is
\begin{equation}\label{id-cond}
 \N_a\equiv\Tr\big[S_aS_a^*\big]<\infty\,.
\end{equation}
Going through the steps~(\ref{fd-ce} --~\ref{fd-res}) in the
present infinite-dimensional context we see that $\E_a$ is finite
if, and only if, $S_a^*h^{1/2}$ extends to a~HS operator, and then
\begin{equation}\label{id-res}
 \E_a=\Tr\big[h^{1/2}S_aS_a^*h^{1/2}\big]\,.
\end{equation}

The quantity $\N_a$ appearing in~(\ref{id-cond}) has a~clear-cut
physical meaning. The results of the Appendix~\ref{bo} show that
if~(\ref{id-cond}) is satisfied, then $\Omega_a\in\D(N)$, where
$N$ is the ``particle'' (excitation) number operator.
A~calculation analogous to that carried out for the energy yields
\begin{equation}\label{id-pnum}
  (\Omega_a,N\Omega_a)=\N_a\,,
\end{equation}
so $\N_a$ is the mean value of the excitation number in the ground
state.

In~the rest of this section we obtain the following criterion for
admissible perturbations. Let $h_a$ be given by~(\ref{id-haform}).
The ground state representations are unitarily equivalent
($\N_a<\infty$) if, and only if,
\begin{equation}\label{id-cdel}
   C_a=\id+\delta_a\,,
\end{equation}
where $\delta_a$ is any operator satisfying conditions
\begin{equation}\label{id-del}
  \delta_a\ \text{is a~HS operator}\,,\quad \id+\delta_a>0\,.
\end{equation}
In~this case we can write in the sense of forms
\begin{equation}\label{id-hdel}
    h_a=h+h^{1/2}\delta_ah^{1/2}\,.
\end{equation}
Moreover, if conditions~(\ref{id-del}) are satisfied, then $\E_a$
is finite if, and only if,
\begin{equation}\label{id-fen}
    \delta_ah^{1/2}\ \text{extends to a~HS operator}\,.
\end{equation}
With the condition~(\ref{id-del}) satisfied one has
\begin{equation}\label{id-n}
    \N_a=\frac{1}{4}\Tr\Big[\frac{\delta_a^2}{\id+\delta_a}\Big]\,,
\end{equation}
and if in addition~(\ref{id-fen}) holds, then
\begin{equation}\label{id-e}
    \E_a=\frac{1}{4}
    \Tr\Big[h^{1/2}\frac{\delta_a^2}{\id+\delta_a}h^{1/2}\Big]\,.
\end{equation}

To prove these assertions note first that equations~(\ref{fd-ts})
remain in force with our assumptions in the present
infinite-dimensional context, and then
\begin{equation}\label{id-ss}
  S_aS_a^*=\frac{1}{4}\frac{(C_a-\id)^2}{C_a}\,.
\end{equation}
If the ground state representations are equivalent, then $S_a^*$
is a~HS operator, so $C_a^{-1/2}(C_a-\id)$ is HS as well.
But~$C_a^{1/2}$ is bounded, therefore also $\delta_a=C_a-\id$ is
HS. The second condition in~(\ref{id-del}) is satisfied by the
positivity of $C_a$. Conversely, suppose that $\delta_a$ satisfies
conditions~(\ref{id-del}). By the first of these conditions
$\delta_a$ has a~purely discrete spectrum with no other
convergence points than zero, and then by the second
$C_a=\id+\delta_a\geq b\,\id$, with $b>0$. Hence $C_a^{-1}$ is
bounded, and $C_a^{-1/2}\delta_a$ is HS, so
 $\N_a=\Tr\big[S_aS_a^*\big]
 =\frac{1}{4}\Tr[\delta_a^2(\id+\delta_a)^{-1}]<\infty$.

If the conditions~(\ref{id-del}) are satisfied, then in
a~completely analogous way one proves the equivalence of the
condition~(\ref{id-fen}) with the finiteness of $\E_a$, and the
equation~(\ref{id-e}).

\setcounter{equation}{0}
\section{Energy density of quantum field}\label{loc}

In~this section we consider the case of a~quantum field, and for
definiteness we take the scalar field (massive or massless) with
standard commutation relations and free evolution. Thus here
$\R=L^2(\mR^3)$, $h=\sqrt{m^2\id-\Delta}$, ($m\geq0$), and we take
for the test function space the largest space $\wh{\Lc}$, as
described in the previous section. Perturbations $h_a$ are assumed
to be in the class defined by \linebreak
Eqs.\,(\ref{id-cdel}~--~\ref{id-fen}) (in fact, a~slight
strengthening of these conditions will be needed). We~show that in
this setting the energy density in the ground states $\Omega_a$ is
well defined as a~tempered distribution, and for the test function
tending to unit function one recovers the energy expectation
value~$\E_a$.

We~assume a~slight strengthening of our assumptions and demand
that for some $\alpha\in(0,1)$ there is:
\begin{equation}\label{loc-loc}
    h^{(1+\alpha)/4}\delta_ah^{(1+\alpha)/4}\quad
    \text{is a~HS operator}\,.
\end{equation}
Note that this statement with $\alpha=0$ is a~consequence of our
earlier assumptions. Indeed, if $\delta_a$ and $\delta_ah^{1/2}$
are HS, then
\begin{multline*}\label{}
    0\leq\Tr(h^{1/4}\delta_ah^{1/4})^2=
    \lim_{n\to\infty}
    \Tr\big[P_{\<0,n\>}(h)(h^{1/4}\delta_ah^{1/4})^2P_{\<0,n\>}(h)\big]\\
    =\lim_{n\to\infty}
    \Tr\big[\delta_ah^{1/2}\delta_ah^{1/2}P_{\<0,n\>}(h)\big]
    =\Tr(\delta_ah^{1/2})^2<\infty\,,
\end{multline*}
where $\{P_F(h)\}$, $F$ a~Borel set in $\mR$, is the spectral
family of $h$.

Loosely speaking, the energy density operator of the scalar field
is determined by the point-splitting procedure and normal ordering
with respect to the vacuum as
\begin{gather}\label{loc-nai}
    H(\ax)=\lim_{\ay\to\ax}:\!H_2(\ax,\ay)\!:\,\equiv
    \lim_{\ay\to\ax}
    \big[H_2(\ax,\ay)-
    (\Omega,H_2(\ax,\ay)\Omega)\big]\,,\\
    H_2(\ax,\ay)=\frac{1}{2}
    \big(P(\ax)P(\ay)+
    \vec{\n}X(\ax)\cdot\vec{\n}X(\ay)+
    m^2X(\ax)X(\ay)\big)\,,\label{loc-h2}
\end{gather}
where $X(u)=\Phi(0,u)$, $P(v)=\Phi(v,0)$ (see~(\ref{cq-ccrf}) and
the preceding remarks), and to get $X(\ax)$ and $P(\ax)$ one sets
formally $v$ and $u$ equal to Dirac delta concentrated at $\ax$.
We~are interested in the energy density
$(\Omega_a,H(\ax)\Omega_a)$ in the ground state~$\Omega_a$.

We~now make this precise. The real Schwartz test function space
$\Sc_\mR$ is contained in $\D_\R(h^{1/2})\cap\D_\R(h^{-1/2})$, so
functions from that space may be used for ``smearing'' both
$X(\ax)$ as $P(\ax)$. Let $w_1,w_2\in\Sc_\mR$. The precise meaning
of~(\ref{loc-h2}) is
\begin{multline}\label{loc-hsm}
    H_2(w_1,w_2)\\=\frac{1}{2}\big(\Phi(w_1,0)\Phi(w_2,0)
    +\Phi(0,\vec{\n}w_1)\cdot\Phi(0,\vec{\n}w_2)+
    m^2\Phi(0,w_1)\Phi(0,w_2)\big)\,.
\end{multline}
To find normal-ordered expectation value
$(\Omega_a,:\!H_2(w_1,w_2)\!:\Omega_a)$ we need to know
\mbox{$(\Omega_a,:\!\Phi(V_1)\Phi(V_2)\!:\Omega_a)$}, where we
assume that $V_i\in\Sc_\mR\oplus\Sc_\mR$. We~recall the
definitions of the representations $\pi$ and $\pi_a$ and their
equivalence relations:
\[
 \Phi(V)=\Phi_0(j(V))\,,\quad \Phi_a(V)=\Phi_0(j_a(V))\,,\quad
 \Omega_a=U_a\Omega\,,\quad U_a\Phi_a(V)U_a^*=\Phi(V)
\]
(Eqs.\,(\ref{cq-repf}), (\ref{fd-def}), (\ref{fd-gr})
and~(\ref{fd-uer}) respectively). Using them one finds
\begin{gather}
  \begin{split}\label{loc-vcf}
    &(\Omega,\Phi(V_1)\Phi(V_2)\Omega)
    =\frac{1}{2}(j(V_1),j(V_2))\\
    &\hspace{60pt}=\frac{1}{2}(h^{1/2}v_1,h^{1/2}v_2)
    +\frac{1}{2}(h^{-1/2}u_1,h^{-1/2}u_2)
    +\frac{i}{2}\sigma(V_1,V_2)\,,
  \end{split}\\
  \begin{split}\label{loc-acf}
    &(\Omega_a,\Phi(V_1)\Phi(V_2)\Omega_a)
    =(\Omega,\Phi_a(V_1)\Phi_a(V_2)\Omega)
    =\frac{1}{2}(j_a(V_1),j_a(V_2))\\
    &\hspace{60pt}=\frac{1}{2}(h_a^{1/2}v_1,h_a^{1/2}v_2)
    +\frac{1}{2}(h_a^{-1/2}u_1,h_a^{-1/2}u_2)
    +\frac{i}{2}\sigma(V_1,V_2)\,,
  \end{split}\\
    (\Omega_a,:\!\Phi(V_1)\Phi(V_2)\!:\Omega_a)
    =\frac{1}{2}(h^{1/2}v_1,\delta_ah^{1/2}v_2)
    -\frac{1}{2}
    \Big(h^{-1/2}u_1,\frac{\delta_a}{\id+\delta_a}h^{-1/2}u_2\Big)
    \,,\label{loc-ancf}
\end{gather}
so
\begin{equation}\label{loc-dis}
    (\Omega_a,:\!H_2(w_1,w_2)\!:\Omega_a)\equiv
    T_a(w_1,w_2)=T_{a1}(w_1,w_2)+T_{a2}(w_1,w_2)\,,
\end{equation}
where
\begin{equation}\label{loc-dis1}
    T_{a1}(w_1,w_2)=\frac{1}{4}
    \Big(h^{1/2}\ov{w_1},\frac{\delta_a^2}{\id+\delta_a}h^{1/2}w_2\Big)\,,
\end{equation}
\begin{multline}\label{loc-dis2}
    T_{a2}(w_1,w_2)=\frac{1}{4}
    \Big(h^{1/2}\ov{w_1},\frac{\delta_a}{\id+\delta_a}h^{1/2}w_2\Big)
    -\frac{m^2}{4}
    \Big(h^{-1/2}\ov{w_1},\frac{\delta_a}{\id+\delta_a}h^{-1/2}w_2\Big)\\
    -\frac{1}{4}
    \Big(h^{-1/2}\vec{\n}\ov{w_1},
    \frac{\delta_a}{\id+\delta_a}h^{-1/2}\vec{\n}w_2\Big)\,.
\end{multline}
We~have added the conjugation sign over $w_1$ on the r.h.\ side to
make the expression linear rather than antilinear also for complex
functions. The $T_{a1}$ part and the first term in $T_{a2}$ result
from splitting
\begin{equation}\label{loc-split}
    \delta_a=\frac{\delta_a^2}{\id+\delta_a}
    +\frac{\delta_a}{\id+\delta_a}\,.
\end{equation}

We~show that:
\begin{itemize}
 \item[(i)] $T_a(w_1,w_2)$ defines a~distribution
$T_a(\ax,\ay)$ on $\Sc(\mR^6)$.
 \item[(ii)] For each $\aet\in\mR^3$ the expression
$T_a(\ak+\aet,\ak-\aet)$ is a~distribution on $\Sc(\mR^3)$, and
for each test function $f$ the function
\begin{equation}\label{loc-diag}
   \aet\to\E_a(\aet,f)=\int
   T_a(\ak+\aet,\ak-\aet) f(\ak)\,d^3\xi
\end{equation}
is continuous and bounded (we use the ``integral'' notation of
distributions). The energy density according to point-splitting
procedure is then the distribution
\begin{equation}\label{loc-enden}
    \E_a(f)\equiv\E_a(\vec{0},f)\,.
\end{equation}
 \item[(iii)] Let $f_\ep(\ak)=f(\ep\ak)$,
 $f(\vec{0})=1$, $f\in\Sc(\mR)$. Then
\begin{equation}\label{loc-inten}
    \lim_{\ep\to0}\,\E_a(f_\ep)=\E_a\,.
\end{equation}
\end{itemize}

Before starting the proof we fix conventions for the Fourier
transforms. \linebreak For~$a,b\in\mR^n$ we set
\begin{equation}\label{loc-ft}
    \hat{f}(b)=(2\pi)^{-n/2}\int
    f(a)e^{-ib\cdot a}d^3a\,,\quad
    \check{f}(a)=\hat{f}(-a)\,.
\end{equation}
 We~consider the $T_{a1}$ and $T_{a2}$ parts
separately. Expressions~(\ref{loc-diag}) and~(\ref{loc-enden}) for
$T_{ai}$ replacing $T_a$ will be denoted $\E_{ai}(\aet,f)$ and
$\E_{ai}(f)$ respectively.

As~$\frac{1}{2}\delta_a(\id+\delta_a)^{-1/2}h^{1/2}$ is a~HS
operator in $L^2(\mR^3)$, it is an integral operator with a~kernel
$k_a(\ax,\ay)\in L^2(\mR^6)$ (see e.g.~\cite{rs}). Thus $T_{a1}$
is obviously a~distribution on $\Sc(\mR^6)$, determined by the
ordinary function
\begin{equation}\label{loc-ker1}
    T_{a1}(\ax,\ay)=\int\ov{k_a(\az,\ax)}k_a(\az,\ay)\,d^3z\,.
\end{equation}
As~for each $\aet$ there is
 $\ov{k_a(\az,\ak+\aet)}k_a(\az,\ak-\aet)\in
 L^1(\mR^6,d^3z\,d^3\xi)$, the distribution $\E_{a1}(\aet,f)$ is
 indeed well defined,
 \begin{equation}\label{loc-distre1}
    \E_{a1}(\aet,f)
    =\int\ov{k_a(\az,\ak+\aet)}f(\ak)k_a(\az,\ak-\aet)\,d^3z\,d^3\xi\,.
\end{equation}
Now, Fourier-transforming $k_a(\az,\ak+\aet)$ and
$f(\ak)k_a(\az,\ak-\aet)$ with respect to $\az$ and $\ak$ one
finds
\begin{equation}\label{loc-e1f}
    \E_{a1}(\aet,f)
    =\frac{1}{(2\pi)^{3/2}}
    \int\ov{\hat{k}_a(\ar,\ap)}\hat{k}_a(\ar,\aq)
    \hat{f}(\ap-\aq)e^{\txt -i\aet\cdot(\ap+\aq)}\,d^3r\,d^3p\,d^3q\,,
\end{equation}
as the integrand on the r.h.\ side is absolutely integrable.
Therefore $\E_{a1}(\aet,f)$ is continuous in $\aet$. For $\aet=0$
we get
\begin{equation}\label{loc-e1}
    \E_{a1}(f)=\int|k_a(\az,\ak)|^2f(\ak)\,d^3zd^3\xi\,.
\end{equation}
As~the function $|k_a(\az,\ak)|^2$ is absolutely integrable, we
see immediately that for $f$ as in (iii) there is
\begin{equation}\label{loc-den1}
    \lim_{\ep\to0}\,\E_{a1}(f_\ep)=
    \int|k_a(\az,\ak)|^2\,d^3z\,d^3\xi
    =\frac{1}{4}
    \Tr\Big[h^{1/2}\frac{\delta_a^2}{\id+\delta_a}h^{1/2}\Big]
    =\E_a\,.
\end{equation}

We~now turn to $T_{a2}$ and take into account our assumption
(\ref{loc-loc}). Using the identity~(\ref{loc-split}) and the fact
that $\delta_a$ and $\delta_ah^{1/2}$ are HS, one finds that an
equivalent formulation of the assumption is that
 $h^{(1+\alpha)/4}\delta_a(\id+\delta_a)^{-1}h^{(1+\alpha)/4}$
is a~HS operator; we denote its kernel in the momentum space by
$l_a(-\ap,\aq)$. We~evaluate $T_{a2}(w_1,w_2)$ in momentum space,
making use of the identity
$\ov{\wh{\ov{w_1}}(\ap)}=\hat{w}_1(-\ap)$:
\begin{equation}\label{loc-t2}
    T_{a2}(w_1,w_2)=\frac{1}{4}\int l_a(\ap,\aq)t(\ap,\aq)
    \hat{w}_1(\ap)\hat{w}_2(\aq)\,d^3p\,d^3q\,,
\end{equation}
where
\begin{equation}\label{loc-t3}
    t(\ap,\aq)
    =\big[(\ap^2+m^2)(\aq^2+m^2)\big]^{(1-\alpha)/8}
    \bigg(1+\frac{\ap\cdot\aq-m^2}
    {\sqrt{(\ap^2+m^2)(\aq^2+m^2)}}\bigg)\,.
\end{equation}
As~$l_a$ is square integrable, and $t$ polynomially bounded,
$T_{a2}$ defines a~distribution $T_{a2}(\ax,\ay)$. Let
$f,g\in\Sc(\mR^3)$. Then
\begin{multline}\label{loc-tgf}
    \int T_{a2}(\ak+\aet,\ak-\aet)f(\ak)g(\aet)\,d^3\xi\,d^3\eta\\
    =2\int l_a(\ar+\as,\ar-\as)\,t(\ar+\as,\ar-\as)
    \hat{f}(2\ar)\hat{g}(2\as)\,d^3r\,d^3s\,,
\end{multline}
where on the l.h.\ side the integral notation is symbolic, but on
the r.h.\ side this is the ordinary integration. Now, one shows
the following estimate
\begin{equation}\label{loc-estim}
    t(\ar+\as,\ar-\as)
    \leq \frac{4|\ar|^2}{(|\ar|^2+|\as|^2+m^2)^{(3+\alpha)/4}}\,,
\end{equation}
(note that $t(\ap,\aq)\geq0$). To prove this it is convenient to
consider the cases\linebreak  $|\ar|^2\geq|\as|^2+m^2$ and
$|\ar|^2<|\as|^2+m^2$ separately. In~the first region one then
uses the obvious bound
$t(\ap,\aq)\leq2\big[(\ap^2+m^2)(\aq^2+m^2)\big]^{(1-\alpha)/8}$,
while in the second one finds that for the given $|\ar|$ and
$|\as|$ the function on the l.h.\ side is the biggest for
$\ar\cdot\as=0$. Using the bound one easily shows that
\begin{equation}\label{loc-est}
    \int[t(\ar+\as,\ar-\as)]^2\,d^3s
    \leq \con\ |\ar|^{4-\alpha}\,.
\end{equation}
Therefore $t(\ar+\as,\ar-\as)\hat{f}(2\ar)\in L^2(\mR^6)$, so
\begin{equation}\label{loc-int6}
    l_a(\ar+\as,\ar-\as)\,t(\ar+\as,\ar-\as)
    \hat{f}(2\ar) \in L^1(\mR^6)\,.
\end{equation}
Using this fact in~(\ref{loc-tgf}) one finds that
\begin{multline}\label{loc-int6f}
    \E_{a2}(\aet,f)=\int T_{a2}(\ak+\aet,\ak-\aet) f(\ak)\,d^3\xi\\
    =\frac{1}{\sqrt{2}\pi^{3/2}}
    \int l_a\Big(\ar+\as,\ar-\as)\,
    t(\ar+\as,\ar-\as)\hat{f}(2\ar)
    e^{\txt -i2\aet\cdot\as}\, d^3r\,d^3s
\end{multline}
indeed defines a~distribution and is continuous in $\aet$. Thus
\begin{equation}\label{loc-enden2}
    \E_{a2}(f)=\frac{1}{\sqrt{2}\pi^{3/2}}
    \int l_a(\ar+\as,\ar-\as)\,t(\ar+\as,\ar-\as)\hat{f}(2\ar)\,
    d^3r\,d^3s\,.
\end{equation}
Using square-integrability of $l_a$ and the estimate
(\ref{loc-est}) we have
\begin{equation}\label{loc-edest}
    |\E_{a2}(f)|^2\leq \con\ \int|\hat{f}(\ar)|^2|\ar|^{4-\alpha}\,d^3r\,.
\end{equation}
For $f$ as in (iii) one easily then finds
\begin{equation}\label{loc-e20}
    \lim_{\ep\to0}\E_{a2}(f_\ep)=0\,,
\end{equation}
which ends the proof of our claims.

In~our calculation of the energy density we have used the standard
definition of the Wick normal ordering. As~an aside, it may be of
interest to mention that this definition has been recently
improved for the cases where the reference state~$\Omega$ depends
on external fields (as e.g.\ in a~fixed curved classical
spacetime). The problem with the usual definition in such cases
is, that the scalar subtraction function depends nonlocally on the
background. This may be remedied, as it turns out, by an
additional subtraction of a~smooth function (in ``Hadamard
states''; see the papers by Hollands and Wald~\cite{hw}, and
Brunetti, Fredenhagen and Verch~\cite{bfv}; for an application to
external field electrodynamics see also~\cite{ma}). This has no
immediate bearing on the discussion in the present work, but may
have applications in related problems with external fields present
from the start (as in a~curved spacetime).

\setcounter{equation}{0}
\section{Remarks on relations with some other\\ appro\-aches}\label{zem}

In~this section we make some remarks on the relation of our
approach to other calculations of Casimir energy. We~shall discuss
a few characteristic examples of local calculations, and next
comment on the ``zero point'' ideology.

First, to make our point on local Casimir energy, we need to
consider a~general situation briefly sketched in the Introduction,
where two representations of local algebras in some open region
$\M_0$ in spacetime are locally quasiequivalent. Suppose we have
two representations $\pi$ and $\tilde{\pi}$ of the algebras of
observables in $\M_0$, acting in Hilbert spaces $\Hc$ and
$\tilde{\Hc}$ respectively. We~assume that the representations are
locally quasiequivalent, but say nothing on their (global)
equivalence. This is the expected state of affairs in many
situations typically considered for Casimir problems. For
instance, for a~scalar field $\M_0$ may be the whole spacetime
outside some $2$-surfaces in $3$-space, $\pi$ the vacuum
representation of the field, and $\tilde{\pi}$ the representation
built on the ground state of the field in presence of the boundary
conditions imposed on the boundaries of $\M_0$. We~choose a~state
in the representation $\tilde{\pi}$, that is a~density operator
$\tilde{\rho}$ in $\tilde{\Hc}$. If the representations are not
equivalent it makes no sense to ask for a~state in the
representation $\pi$ which gives the same expectation values as
$\tilde{\rho}$ for all observables. However, the local
quasiequivalence tells us that if we restrict attention to an open
subset $\Oc$ with a~compact closure contained in $\M_0$, then
there exists a~density operator $\rho_\Oc$ in $\Hc$ such that
\begin{equation}\label{zem-loc}
 \Tr[\tilde{\rho}\tilde{\pi}(A)]=\Tr[\rho_\Oc\pi(A)]\quad
 \text{for}\quad A\ \text{in}\ \Oc\,.
\end{equation}
The local energy density is not one of the fundamental local
observables $A$, but it may be locally built with the use of them.
In~the sequel we restrict attention to the scalar field and
construct local energy density as in~(\ref{loc-nai}). Thus given
a~state $\tilde{\rho}$ the Casimir energy density in $\Oc$
according to the views we follow in this paper is
\begin{equation}
 \tilde{\E}(\vec{x})=\Tr[\rho_\Oc H(\vec{x})]
\end{equation}
(expectation value of a~fixed, \emph{free} field energy density
operator). Let $\tilde{\rho}$ be, for simplicity, the projection
operator onto the unit vector $\tilde{\W}$. Then using
(\ref{loc-nai}) and~(\ref{zem-loc}) we can write for the Casimir
energy at a~given time $t=0$:
\begin{equation}
 \begin{split}
 \tilde{\E}(\vec{x})=&\frac{1}{2}\lim_{\ax'\to\ax}\Big\{
 \big(\tilde{\W},\big[\tilde{P}(\ax)\tilde{P}(\ax')+
 \vec{\n}\tilde{X}(\ax)\cdot\vec{\n}\tilde{X}(\ax')+
 m^2\tilde{X}(\ax)\tilde{X}(\ax')\big]\tilde{\W}\big)\\
 &-\big(\W,\big[P(\ax)P(\ax')+
    \vec{\n}X(\ax)\cdot\vec{\n}X(\ax')+
    m^2X(\ax)X(\ax')\big]\W\big)\Big\}\,,
 \end{split}
\end{equation}
where $X(u)=\Phi(0,u)$, $P(v)=\Phi(v,0)$,
$\tilde{X}(u)=\tilde{\Phi}(0,u)$,
$\tilde{P}(v)=\tilde{\Phi}(v,0)$, and $\Phi$ and $\tilde{\Phi}$
are operators representing the field under $\pi$ and $\tilde{\pi}$
respectively. Recall that in the case discussed in Section
\ref{loc} there is $\tilde{\Hc}=\Hc$, $\tilde{\W}=\W$,
$\tilde{\Phi}=\Phi_a$, and one recovers the formula obtained at
the beginning of that section. More generally, let in each of the
representations a~different time evolution be given by unitary
operators: free evolution $U(t)$ and evolution influenced by
background $\tilde{U}(t)$ respectively, and denote
\begin{equation}
 \begin{split}
 &X_t(\ax)=U(t)X(\ax)U(t)^*\equiv\varphi(t,\ax)\,,\quad
 P_t(\ax)=U(t)P(\ax)U(t)^*\,,\\
 &\tilde{X}_t(\ax)=U(t)\tilde{X}(\ax)U(t)^*
 \equiv\tilde{\varphi}(t,\ax)\,,\quad
 \tilde{P}_t(\ax)=U(t)\tilde{P}(\ax)U(t)^*\,.
 \end{split}
\end{equation}
If for both evolutions there is
$P_t(\ax)=\partial\varphi(t,\ax)/\partial t$,
$\tilde{P}_t(\ax)=\partial\tilde{\varphi}(t,\ax)/\partial t$, then
one can write the last formula for $\tilde{\E}(\ax)$ at $t=0$ as
\begin{equation}\label{zem-kay}
 \tilde{\E}(\ax)=\frac{1}{2}\lim_{\substack{t,t'\to0\\ \ax'\to\ax}}
 \big\{\partial_t\partial_{t'}+\vec{\n}\cdot\vec{\n}'+m^2\big\}
 \big\{(\tilde{\W},\tilde{\varphi}(t,\ax)
 \tilde{\varphi}(t',\ax')\tilde{\W})
 -(\W,\varphi(t,\ax)\varphi(t',\ax')\W)\big\}
\end{equation}
This formula was derived along similar lines by Kay~\cite{kay} in
the context of the free field in a~locally flat spacetime with
nontrivial topology. There are no boundaries in that case, but the
net of local algebras of observables in this spacetime differs
from that in the globally flat Minkowski spacetime, so the notion
of a~global Casimir energy in the sense we use here has no
application.

In~the context of electromagnetic field bounded by conductors in
Minkowski space the opinion similar to ours, that one should
compare expectation values of the fixed free field energy density,
was expressed by Scharf and Wreszinski~\cite{sw}. Consider
a~massless scalar field analogy of the setting. Then $\M_0$ is the
spacetime region outside boundaries. Put $m=0$ in the last
formula, use the translational symmetry of the two-point functions
(which enables the replacement $\vec{\n}'\rightarrow-\vec{\n}$),
and the wave equation, which both correlation functions satisfy
outside the boundaries. This leaves us with
\begin{equation}
 \tilde{\E}(\ax)=\frac{1}{2}\lim_{\substack{t,t'\to0\\ \ax'\to\ax}}
 \big\{\partial_t\partial_{t'}+\partial_t^2\big\}
 \big\{(\tilde{\W},\tilde{\varphi}(t,\ax)
 \tilde{\varphi}(t',\ax')\tilde{\W})
 -(\W,\varphi(t,\ax)\varphi(t',\ax')\W)\big\}\,,
\end{equation}
which is the formula used in~\cite{sw}. No global energy density
may be obtained in this way (if not by an \emph{ad hoc}
regularization of the infinities in the density) due to the
algebraic problems explained earlier.

Next, we want to comment on the ``Green function'' method.
In~papers following this method one usually states that the
(local) Casimir energy is the difference between the energy ``in
the vacuum state with the barriers present and with them absent''
(see e.g.\,\cite{dc}), with no further explanation on what energy
is meant. Staying with the scalar field as our example, one then
uses with not much comment a~formula similar to~(\ref{zem-kay}),
in which, however, the products of fields are replaced by
time-ordered products. This brings no change of the result in this
simple case, but in general has to be justified. As~long as
outside the barriers the field follows the same local equation
(the distinct time evolutions agree locally), the ambiguity as to
what energy is meant does not show up. However, this does not
matter only because for sharp boundaries one cannot determine the
global energy anyway. And in fact, if the barriers are replaced by
external fields one has to make it clear what is being calculated.
An example of such calculation is attempted in~\cite{pmg}, where
one of the sections treats on the quantum Dirac field in an
external classical electromagnetic field. The authors' intention
apparently is to compare the energy of the Dirac field itself, so
they keep the free field energy expression. However, they take
over the form of this expression containing time derivatives, and
to eliminate them they use different field equations in the two
cases, which spoils the original intention (remember that the
Dirac equation is first order, so the time derivative of the field
is not an independent initial value variable).

Another example of the external field calculations is to be found
in~\cite{gj}. Here the authors with the intention of finding the
global Casimir energy explicitly compare expectation values of two
different energy operators: energy of the field with the
interaction terms included and the energy as given by the free
field theory, in the ground states of the two respective
evolutions. In~our opinion this is one of the reasons for the
appearance of infinities in their expressions, which are
eliminated by adding ``counterterms'' to the model which does not
need them (except for trivial normal ordering of quadratic
observables). We~note, moreover, that it does not follow from the
smoothness of the external field alone that the ground state
representations, with the external field present or not, are
equivalent globally.

In~the rest of this section we try to understand, from the point
of view of the formalism presented in this paper, how ``zero
point'' expressions may arise in the context of Casimir effect for
quantum fields. In~our opinion their appearance is a~consequence
of unjustified manipulations. Accordingly, the equations and
transformations to be found below are not to be taken at face
value. We~indicate this by putting a~dot over the equality sign.

The ``zero point'' expression for Casimir energy has the form
\begin{equation}\label{zem-zem}
    \E_a^{\mathrm{z.p.}}\doteq \frac{1}{2}\sum_k\omega_{ak}-
    \frac{1}{2}\sum_k\omega_k\,,
\end{equation}
where $\omega_k$ and $\omega_{ak}$ are appropriately discretized
frequencies of free and perturbed field respectively. In~our
language this would be
\begin{equation}\label{zem-zemtr}
    \E_a^{\mathrm{z.p.}}\doteq \frac{1}{2}\Tr(h_a-h)\,,
\end{equation}
which usually is meaningless, but is then ``regularized'' to
squeeze a~finite result. We~show how this expression may arise.

We~have shown in the previous section that $\E_a$ is a~limit of
the energy density distribution value for the test function
tending to one, see Eq.\,(\ref{loc-inten}). Also, it turned out
that in this limit only the part $\E_{a1}(f)$,
Eq.\,(\ref{loc-e1}), of the density distribution contributes. Thus
we can use Eq.\,(\ref{loc-den1}) to calculate the total energy.
Distribution $\E_{a1}(f)$ is determined by part $T_{a1}$,
Eq.\,(\ref{loc-dis1}), of $(\Omega_a,:\!H_2(w_1,w_2)\!:\Omega_a)$.
If we do not pay due attention to domains we can rewrite
Eq.\,(\ref{loc-dis1}) by expressing it in terms of $h$ and $h_a$
instead of $h$ and $\delta_a$. The result is
\begin{equation}\label{zem-t1}
    T_{a1}(w_1,w_2)
    \doteq\frac{1}{4}\int \<\ax|h_a-h+h(h_a^{-1}-h^{-1})h|\ay\>
    w_1(\ax)w_2(\ay)\,d^3x\,d^3y\,,
\end{equation}
which implies
\begin{multline}\label{zem-e}
    \E_a\doteq\frac{1}{4}\int
    \<\ax|h_a-h+h(h_a^{-1}-h^{-1})h|\ax\>\,d^3x\\
    \doteq\frac{1}{4}\Tr\big[h_a-h+h(h_a^{-1}-h^{-1})h\big]\,.
\end{multline}
 Let us now, again ignoring difficulties, apply this
to the case of sharp boundaries, where $h_a^2=-\Delta_{B(a)}$ with
appropriate boundary conditions $B(a)$. Suppose that the support
of $w_1$ and $w_2$ stays outside the boundaries. Then
$h_a^2w_i=h^2w_i$, and we have
\begin{equation}\label{zem-hah}
    (\ov{w_1},(h_a-h)w_2)
    \doteq\frac{1}{2}(\ov{w_1},(h_a^{-1}-h^{-1})h^2w_2)
    +\frac{1}{2}(\ov{w_1},h^2(h_a^{-1}-h^{-1})w_2)\,,
\end{equation}
(written in two terms only for symmetry reasons), or, for $\ax$
and $\ay$ outside the boundary,
\begin{equation}\label{zem-hah1}
    \<\ax|h_a-h|\ay\>
    \doteq\frac{1}{2}\<\ax|(h_a^{-1}-h^{-1})h^2
    +h^2(h_a^{-1}-h^{-1})|\ay\>\,.
\end{equation}
This needs regularization on the boundaries. Assuming some form of
it one writes
\begin{multline}\label{zem-hah2}
    \Tr(h_a-h)\doteq\frac{1}{2}\int\<\ax|(h_a^{-1}-h^{-1})h^2
    +h^2(h_a^{-1}-h^{-1})|\ax\>\,d^3x\\
    \doteq\frac{1}{2}\Tr\big[(h_a^{-1}-h^{-1})h^2
    +h^2(h_a^{-1}-h^{-1})\big]\,.
\end{multline}
Suppose that the regularization used cuts high momenta, so as to
allow one to change the order of operators under the trace sign.
Then
\begin{equation}\label{zem-hah3}
    \Tr(h_a-h)\doteq\Tr\big[h(h_a^{-1}-h^{-1})h\big]\,.
\end{equation}
Using this in~(\ref{zem-e}) one arrives at~(\ref{zem-zemtr}).

\setcounter{equation}{0}
\appendix
\section{Appendix. Fock space operators and Bo\-go\-liu\-bov
transformations}\label{appA}

In~the appendix we give a~brief review of some known properties of
Fock space operators which are needed in the main text. The main
sources of reference for Section~\ref{we} are books~\cite{br}
(vol.\,II) and~\cite{rs}. The content of Sections~\ref{st} and
\ref{bo} is a~rather common knowledge. Precise original proofs of
the criterions of equivalence of representation use rather more
advanced and less common techniques~\cite{sdv}, so we think
a~simple proof with the use of creation/annihilation operators is
worth presenting in~\ref{pr}. (The results on equivalence have
been later generalized, in the widest form in~\cite{ay}).

\subsection{Weyl system in a~Fock space}\label{we}

Let $\Hc$ be the symmetric Fock space based on the ``one-particle
(excitation) space'' $\K$, i.e.
\begin{equation}\label{we-fock}
  \Hc=\bigoplus_{n=0}^\infty \Hc_n\,,\quad \Hc_0=\mathbb{C}\,,\quad
  \Hc_n=\Sc(\underbrace{\K\otimes\ldots\otimes\K}_{n\ \text{times}})\ \
  (n\geq 1)\,,
\end{equation}
where $\Sc$ is the symmetrization projection operator. The scalar
product in $\Hc$ will be denoted by $(.\,,.)$, the ``Fock vacuum''
vector by $\Omega$, and the particle (excitation) number operator
by $N$. On the domain \mbox{$\D(N^{1/2})$} the annihilation and
creation operators are defined in the usual way: for each $f\in\K$
and $\psi,\chi\in\D(N^{1/2})$ one sets
\begin{equation}\label{we-ac}
  a^*(f)\psi=\Sc(f\otimes\sqrt{N+1}\,\psi)\,,\qquad
  (\chi,a(f)\psi)=(a^*(f)\chi,\psi)\,,
\end{equation}
and shows that
\begin{equation}\label{we-acb}
  \|a^\#(f)\psi\|\leq\|f\|\,\|(N+1)^{1/2}\psi\|\,,\quad
  a^\#(f)=a(f)\ \text{or}\ a^*(f)\,,
\end{equation}
and for $\varphi\in\D(N)$
\begin{equation}\label{we-acom}
 [a(f),a^*(g)]\varphi=(f,g)\varphi\,.
\end{equation}
 Operators $a(f)$ and $a^*(f)$ are respectively
antilinear and linear in $f$.

Let $\Hc_f$ be the finite-excitation subspace (dense in $\Hc$),
i.e.
\begin{equation}\label{we-fockf}
  \Hc_f=\bigcup_{k=0}^\infty\ \bigoplus_{n=0}^k\Hc_n\,.
\end{equation}
Operators $\FF(f)$ are defined in the following way. One initially
sets
\begin{equation}\label{we-fi}
  \FF(f)\psi=\frac{1}{\sqrt{2}}\big(a(f)+a^*(f)\big)\psi\quad
  \text{for}\quad  \psi\in\D(N^{1/2})\,.
\end{equation}
Using the bounds~(\ref{we-acb}) one shows that these operators are
essentially selfadjoint on $\Hc_f$, so their closures $\FF(f)$ are
selfadjoint. For $\psi\in\D(N^{1/2})$, $\varphi\in\D(N)$,
$f,g,f_k\in\K$ and real $\alpha,\beta$ one has
\begin{gather}
  \FF(\alpha f+\beta g)\psi=\alpha\FF(f)\psi+\beta\FF(g)\psi\,,
  \label{we-fil}\\
  a(f)\psi=\frac{1}{\sqrt{2}}\big(\FF(f)+i\FF(if)\big)\psi\,,\quad
  a^*(f)\psi=\frac{1}{\sqrt{2}}\big(\FF(f)-i\FF(if)\big)\psi\,,
  \label{we-afi}\\
  \text{if}\ \ \|f_k-f\|\to 0\ \ \text{then}\ \
  \|\FF(f_k)\psi-\FF(f)\psi\|\to 0\quad
  (k\to\infty)\,,\label{we-lfi}\\
  [\FF(f),\FF(g)]\varphi=i\Ip(f,g)\varphi\,.\label{we-fic}
\end{gather}
Using these relations one shows that the Weyl operators defined by
\begin{equation}\label{we-de}
  \WF(f)=e^{\textstyle i\FF(f)}
\end{equation}
have the following properties
\begin{gather}
 \WF(f)\WF(g)=e^{-\frac{i}{2}\Ip(f,g)}\WF(f+g)\,,\
 \WF(f)^*=\WF(-f)\,,\ \WF(0)=\id\,;\label{we-we}\\
 \text{the set}\ \{\WF(f)\mid f\in\K\}\ \text{is irreducible}\,;
 \label{we-weir}\\
 (\Omega,\WF(f)\Omega)=e^{\textstyle
 -\frac{1}{4}\|f\|^2}\,;\label{we-state}\\
 \text{if}\ \ \|f_k-f\|\to 0\ \ \text{then}\ \
  \|\WF(f_k)\psi-\WF(f)\psi\|\to 0\quad
  (k\to\infty)\,,\ \psi\in\Hc\,.\label{we-lwe}
\end{gather}

Let $U$ be a~unitary operator in $\K$. One defines a~unitary
operator $\Gamma(U)$ in $\Hc$ by
\begin{equation}\label{we-ghw}
  \Gamma(U)\WF(f)\Omega=\WF(Uf)\Omega\,,
\end{equation}
which implies
\begin{equation}\label{we-ghi}
  \Gamma(U)\WF(f)\Gamma(U)^*=\WF(Uf)\,.
\end{equation}
It is then easy to show that
\begin{equation}\label{we-gh}
  \Gamma(U):\Hc_n\mapsto\Hc_n\,,\quad
  \Gamma(U)\big|_{\Hc_n}=
  \underbrace{U\otimes\ldots\otimes U}_{n\ \text{times}}\,.
\end{equation}
Let now $h$ be a~selfadjoint operator in $\K$. Then
$\Gamma(e^{ith})$ is a~one-parameter group of unitary operators.
The generator of this group, denoted $d\Gamma(h)$, is
a~selfadjoint operator,
\begin{equation}\label{we-dgh}
  \Gamma(e^{ith})=\exp(itd\Gamma(h))\,.
\end{equation}
Let $\D_h$ be any domain of essential selfadjointness of $h$ and
denote
\begin{equation}\label{we-ddgh}
  \D_{d\Gamma(h)}=\bigcup_{k=0}^\infty\ \bigoplus_{n=0}^k
  \Sc(\underbrace{\D_h\odot\ldots\odot\D_h}_{n\ \text{times}})\,,
\end{equation}
which means that $\D_{d\Gamma(h)}$ is formed by finite linear
combinations of symmetrized products of vectors from $\D_h$. One
shows that
\begin{equation}\label{we-esdg}
  d\Gamma(h)\big|_{\Sc(\D_h\odot\ldots\odot\D_h)}
  =h\otimes\id\otimes\ldots\otimes\id+\ldots+
  \id\otimes\ldots\otimes\id\otimes h\,,
\end{equation}
and that $d\Gamma(h)$ is essentially selfadjoint on
$\D_{d\Gamma(h)}$. We~assume now that $h$ is a~nonnegative
operator. Then $d\Gamma(h)$ is also nonnegative and has the
following representation in terms of quadratic forms.
As~$(a(f))^*$ is densely defined (its domain contains
$\D(N^{1/2})$), the annihilation operator $a(f)$ is closable, we
denote its closure by $\bar{a}(f)$. Let $\{f_i\}$ be any
orthonormal basis of $\K$ formed of vectors in $\D(h^{1/2})$, and
denote
\begin{equation}\label{we-qad}
  Q(q)=\{\psi\in\Hc\mid \psi\in\bigcap_{i}\D(\bar{a}(h^{1/2}f_i))\ \text{and}\
  \sum_i\|\bar{a}(h^{1/2}f_i)\psi\|^2<\infty\}\,.
\end{equation}
One shows that the following form on $Q(q)$ is closed
\begin{equation}\label{we-qf}
  q(\psi,\chi)=\sum_i(\bar{a}(h^{1/2}f_i)\psi,\bar{a}(h^{1/2}f_i)\chi)\,.
\end{equation}
It is easy to check by direct calculation that the restriction of
this form to $\D_{d\Gamma(h)}$ gives
\begin{equation}\label{we-qfo}
  q(\psi,\chi)=(\psi,d\Gamma(h)\chi)\,.
\end{equation}
As~$\D_{d\Gamma(h)}$ is a~core of $d\Gamma(h)$, the unique
selfadjoint operator defined by the form~$q$ is identical with
$d\Gamma(h)$. Thus
\begin{equation}\label{we-qh}
 Q(q)=\D({d\Gamma(h)}^{1/2})\quad \text{and}\quad
 q(\psi,\chi)=({d\Gamma(h)}^{1/2}\psi,{d\Gamma(h)}^{1/2}\chi)\,.
\end{equation}
In~particular, for all $\psi,\chi\in\D(d\Gamma(h))$ identity
(\ref{we-qfo}) holds.

We~note that the particle number operator may be represented as
a~special case of this construction,
\begin{equation}\label{we-part}
 N=d\Gamma(\id)\,.
\end{equation}

\subsection{Symplectic transformations of
$(\K,\Ip(.\,,.))$}\label{st}

Hilbert space $\K$, as a~real vector space, is a~symplectic space
with the form $\Ip(.\,,.)$. Its real-liner, bijective
transformation $L$ is a~symplectic transformation if for all
$f,g\in\K$
\begin{equation}\label{st-st}
  \Ip(Lf,Lg)=\Ip(f,g)\,.
\end{equation}
The inverse transformation is then also a~symplectic
transformation satisfying the same condition. Substituting $f\to
L^{-1}f$ in~(\ref{st-st}) one has
\begin{equation}\label{st-ti}
  \Ip(f,Lg)=\Ip(L^{-1}f,g)\,.
\end{equation}
One defines operators on $\K$:
\begin{gather}
  T=\tfrac{1}{2}(L-iLi)\,,\ \ S=\tfrac{1}{2}(L+iLi)\,,\ \
  L=T+S\,,\label{st-ts}\\
  T'=\tfrac{1}{2}(L^{-1}-iL^{-1}i)\,,\ \
  S'=\tfrac{1}{2}(L^{-1}+iL^{-1}i)\,,\ \
  L^{-1}=T'+S'\,.\label{st-tspr}
\end{gather}
Operators $T$ and $T'$ are complex-linear, while $S$ and $S'$ are
complex-antilinear. Using their definitions and the relation
(\ref{st-ti}) it is easy to show that operators in the two pairs
$T',T$ and $S',-S$ are mutually adjoint, so $T=T^{**}$ and
$S=S^{**}$. Thus both operators are everywhere defined and closed,
so they are bounded. Separating the identities $L^{-1}L=\id$ and
$LL^{-1}=\id$ into linear and antilinear parts one gets
\begin{gather}
 T^*T=S^*S+\id\,,\quad T^*S=S^*T\,,\label{st-tst}\\
 TT^*=SS^*+\id\,,\quad TS^*=ST^*\,.\label{st-tts}
\end{gather}
Conversely, if the operators $T$ and $S$ satisfy the above
relations on the whole Hilbert space $\K$, they are bounded and
define a~symplectic transformation\linebreak $L=T+S$. Furthermore,
if the relations are satisfied, then $T$ is a~bijection of $\K$
onto $\K$. Thus if $T=U_T|T|$ is its unique polar decomposition,
then $U_T$ is a~unitary operator. We~set $S=U_TR$. It follows then
from the first equalities in~(\ref{st-tst}) and~(\ref{st-tts})
that $R^*R=RR^*$. If $R=K|S|$ is the unique polar decomposition of
$R$, then this condition is equivalent to $K|S|=|S|K$, so $K$ is
a~partial anti-isometry of $(\Ker|S|)^\bot$ onto itself. From the
first relation in~(\ref{st-tst}) $|T|^2=\id+|S|^2$, so $K|T|=|T|K$
as well. The second relation in~(\ref{st-tst}) then gives $K^*=K$.
We~summarize the results:
\begin{gather}
 T=U_T(\id+|S|^2)^{1/2}\,,\quad S=U_T|S|K\,,\quad [|S|,K]=0\,,
 \label{st-pd}\\
 |S|\ \text{is bounded}\,,\ U_T\ \text{is unitary}\,,\
 K\ \text{is a~conjugation on}\ (\Ker|S|)^\bot\,.\label{st-pdc}
\end{gather}
Conversely, if these conditions are satisfied, then $T$ and $S$
satisfy conditions~(\ref{st-tst}) and~(\ref{st-tts}), and
determine a~symplectic transformation by
\begin{equation}\label{st-linl}
  L=T+S\,,\qquad L^{-1}=T^*-S^*\,.
\end{equation}

If $|S|$ has no continuous spectrum, then it follows from the
above relations that its orthonormal basis of eigenvectors may be
chosen such that
\begin{equation}\label{st-ei}
  |S|f_i=\lambda_if_i\,,\quad Kf_i=f_i\,.
\end{equation}

\subsection{Bogoliubov transformations in a~Fock space}\label{bo}

With the notation of the foregoing subsections let $L=T+S$ be
a~symplectic transformation of the space $(\K,\Ip(.,.))$, and let
us denote
\begin{equation}\label{bo-de}
  \WF_L(f)=\WF(Lf)\,.
\end{equation}
It is easy to show that these new operators also satisfy the Weyl
relations~(\ref{we-we}). The transformation
$\WF(f)\mapsto\WF_L(f)$ is called a~Bogoliubov transformation. Its
equivalent form is
\begin{equation}\label{bo-fi}
  \FF(f)\mapsto\FF_L(f)=\FF(Lf)\,,\quad
  \WF_L(f)=e^{\textstyle  i\FF_L(f)}\,.
\end{equation}
For $\psi\in\D(\FF(Lf))\cap\D(\FF(Lif))$ one defines
\begin{equation}\label{bo-al}
  a_L(f)\psi=\frac{1}{\sqrt{2}}\big(\FF_L(f)
  +i\FF_L(if)\big)\psi\,,\quad
  a_L^*(f)\psi=\frac{1}{\sqrt{2}}\big(\FF_L(f)
  -i\FF_L(if)\big)\psi\,,
\end{equation}
and shows by a~simple calculation that for $\psi\in\D(N^{1/2})$:
\begin{equation}\label{bo-abo}
  a_L(f)\psi=a(Tf)\psi+a^*(Sf)\psi\,,\quad
  a_L^*(f)\psi=a^*(Tf)\psi+a(Sf)\psi\,.
\end{equation}
Then using Eqs.\,(\ref{st-tst}) one also finds for
$\psi\in\D(N^{1/2})$
\begin{equation}\label{bo-aboin}
  a(f)\psi=a_L(T^*f)\psi-a_L^*(S^*f)\psi\,,\quad
  a^*(f)\psi=a_L^*(T^*f)\psi-a_L(S^*f)\psi\,,
\end{equation}
and for $\varphi\in\D(N)$
\begin{equation}\label{bo-acom}
 [a_L(f),a^*_L(g)]\varphi=(f,g)\varphi\,.
\end{equation}

One says that the Bogoliubov transformation is implementable in
$\Hc$ if there exists a~unitary operator $U_L$ such that either of
the following (and then both) conditions hold
\begin{equation}\label{bo-im}
  \WF_L(f)=U_L\WF(f)U_L^*\,,\quad
  \FF_L(f)=U_L\FF(f)U_L^*\,,\quad  f\in\K\,.
\end{equation}

The necessary and sufficient condition for the implementability of
the Bogoliubov transformation is that $S$ be a~Hilbert-Schmidt
operator, i.e.
\begin{equation}\label{bo-hs}
  \Tr\big[S^*S\big]<\infty\,.
\end{equation}
If the condition is satisfied, then there exists a~unique, up to
a~phase factor, normalized vector $\Omega_L$ satisfying the
conditions
\begin{equation}\label{bo-an}
  a_L(f)\Omega_L=0\,,\qquad f\in\K\,.
\end{equation}
Moreover, one has
\begin{equation}\label{bo-nv}
  \Omega_L\in\bigcap_{l=1}^\infty\D(N^{l/2})\,.
\end{equation}
Equations
\begin{equation}\label{bo-dei}
  U_L\,a^*(f_1)\ldots a^*(f_k)\Omega=a^*_L(f_1)\ldots
  a^*_L(f_k)\Omega_L\,,\quad k=0,1,\ldots\,,
\end{equation}
with arbitrary test vectors $f_i$, define the unique (up to
a~phase factor) unitary operator $U_L$ implementing the Bogoliubov
transformation. For the completeness we sketch a~simple proof of
these statements in the next subsection.

A~slight generalization of the above results is needed in the main
text. Let $\J$ and $\J'$ be real subspaces of $\K$, dense in $\K$,
and let $L:\J\mapsto\J'$ be a~bijective symplectic transformation
(i.e. a~real-linear transformation satisfying Eq.\,(\ref{st-st})
for $f\in\J$). Suppose that there exists a~unitary operator $U_L$
such that
\begin{equation}\label{bo-impg}
 \WF(Lf)=U_L\WF(f)U_L^*\,, \quad f\in\J\,.
\end{equation}
Then $L$ and $L^{-1}$ extend to bounded symplectic transformations
on $\K$, and \linebreak Eq.\,(\ref{bo-im}) is satisfied for all
$f\in\K$.

Indeed, suppose that~(\ref{bo-impg}) is fulfilled. Then using
Eq.\,(\ref{we-state}) one finds
\begin{equation}
 e^{\txt -\frac{1}{4}\|Lf\|^2}=(U_L^*\Omega,\WF(f)U_L^*\Omega)\,,
 \quad f\in\J\,,
\end{equation}
which shows that $L$ is a~continuous transformation on its domain
(if $f_n\to 0$, then by Eq.\,(\ref{we-lwe}) also $Lf_n\to 0$).
Thus $L$ extends by continuity to a~bounded operator on $\K$. From
Eq.\,(\ref{bo-impg}) we have $U_L^*\WF(f)U_L=\WF(L^{-1}f)$ for
$f\in\J'$, thus similar reasoning shows that the extension of $L$
is a~bijective symplectic transformation of $\K$ onto itself.
Equation~(\ref{bo-impg}) now extends by~(\ref{we-lwe}) to all
$f\in\K$.

\subsection{Proof of the statements~(\ref{bo-hs} -- \ref{bo-dei})}
\label{pr}

Let the Bogoliubov transformation be implemented as in
Eq.\,(\ref{bo-im}). For each pair of vectors
$\psi,\varphi\in\D(N^{1/2})$ one has then
 $(a_L^*(f)\psi,U_L\varphi)=(\psi,U_La(f)\varphi)$. We~substitute here
\mbox{$\varphi=\Omega$}, $f=T^{-1}g$, and use Eq.\,(\ref{bo-abo}).
This yields
\begin{equation}\label{pr-po}
 (a^*(g)\psi,\Omega_L)=-(a(ST^{-1}g)\psi,\Omega_L)\,,
\end{equation}
where $\Omega_L=U_L\Omega$. Substituting here for $\psi$ all
vectors of the form $a^*(g_1)\ldots a^*(g_k)\Omega$ for
$k=0,1,\ldots$ recursively, it is easy to see that
$(\Omega,\Omega_L)$ cannot vanish, as otherwise $\Omega_L$ would
be orthogonal to the whole Hilbert space. Let now $\{f_i\}$ be an
orthonormal basis and put in~(\ref{pr-po}) $g=f_i$ and
$\psi=a^*(f_j)\Omega$, which gives
\begin{equation}\label{pr-str}
  (a^*(f_i)a^*(f_j)\Omega,\Omega_L)
  =-(ST^{-1}f_i,f_j)(\Omega,\Omega_L)\,.
\end{equation}
Take the sum over $i,j$ of the absolute values squared of both
sides of this equation. On the l.h.\ side one then gets a~quantity
smaller or equal $2\|\Omega_L\|^2$, so $ST^{-1}$ is
a~Hilbert-Schmidt operator. As~$T$ is a~bounded operator, the same
is true for $S$.

Conversely, let now $S$ be a~HS operator, so there exists an
orthonormal basis $\{f_i\}$ satisfying~(\ref{st-ei}), and denote
$g_i=U_Tf_i$, which defines another orthonormal basis. Then
\begin{equation}\label{pr-fs}
 \begin{split}
 &Sf_i=\lambda_ig_i\,,\quad Tf_i=\sqrt{\lambda_i^2+1}g_i\,,\quad
 \text{so}\quad
 ST^{-1}g_i=\frac{\lambda_i}{\sqrt{\lambda_i^2+1}}g_i\,,\\
 &\hspace{99pt}\text{where}\quad \sum_{i=1}^\infty\lambda_i^2<\infty\,.
 \end{split}
\end{equation}
We~look for a~vector $\Omega_L$ which lies in the domain of all
operators $a_L(f)$ and satisfies Eq.\,(\ref{bo-an}). If such
vector exists, then it must satisfy Eq.\,(\ref{pr-po}) for all
possible $g$ and $\psi$. It is sufficient to substitute for $g$
all basis vectors $g_i$ and for $\psi$ all vectors from the basis
of the particle number representation $\{|n_1,n_2,\ldots\>\}$ with
profiles $g_1,g_2,\ldots$. This gives the recurrent conditions
\begin{equation}\label{pr-rec}
  \<n_1\ldots n_i+1\ldots|\Omega_L\>
  =-\frac{\lambda_i}{\sqrt{\lambda_i^2+1}}\sqrt{\frac{n_i}{n_i+1}}
  \<n_1\ldots n_i-1\ldots|\Omega_L\>\,.,
\end{equation}
which are solved for numbers $\<n_1n_2\ldots|\Omega_L\>$ uniquely
up to a~common constant factor $c$ by
\begin{equation}\label{pr-eta}
 \begin{split}
 \<n_1n_2\ldots|\Omega_L\>&=0\qquad \text{if not all $n_i$ are even}\,,\\
 \<2m_1\,2m_2\ldots|\Omega_L\>&=c\,
 \prod_{i=1}^\infty\bigg(-\frac{\lambda_i}{\sqrt{\lambda_i^2+1}}\bigg)^{m_i}\,
 \sqrt{\frac{(2m_i-1)!!}{(2m_i)!!}}\,,
 \end{split}
\end{equation}
Using these explicit expressions one finds that for each
$l=0,1,\ldots$ the following sum converges:
\begin{multline}\label{pr-nl}
 \sum_{n_1,n_2,\ldots}\bigg(\sum_{i=1}^\infty n_i\bigg)^l
 |\<n_1n_2\ldots|\Omega_L\>|^2\\
 \leq|c|^22^l
 \sum_{m_1,m_2,\ldots}\bigg(\sum_{i=1}^\infty m_i\bigg)^l
 \prod_{j=1}^\infty
 \bigg(\frac{\lambda_j^2}{1+\lambda_j^2}\bigg)^{m_j}<\infty\,.
\end{multline}
The first inequality is obvious, while the second bound will be
shown below. For $l=0$ the bound shows that the coefficients
$\<n_1n_2\ldots|\Omega_L\>$ indeed define a~vector~$\Omega_L$
solving the conditions~(\ref{pr-rec}). The bounds for $l\in\mN$
show that this vector is in the domain of all operators $N^{l/2}$.
In~particular, $\Omega_L$ is in the domain of all operators
$a_L(f)$. This completes the proof of the existence and uniqueness
(up to a~phase) of a~normalized vector solving equation
(\ref{bo-an}), and of the property~(\ref{bo-nv}). Statements about
the operator $U_L$ are now easily proved with the use of
commutation relations, and the irreducibility of the Weyl system.

To prove the missing step in Eq.\,(\ref{pr-nl}) we denote
$q_i=\lambda_i^2(1+\lambda_i^2)^{-1}$. From the finiteness of the
sum $\dsp\sum_{i=1}^\infty\lambda_i^2$ it follows that also the
following expressions converge:
\begin{gather*}
 p_l\equiv\sum_{i=1}^\infty\bigg(\frac{q_i}{1-q_i}\bigg)^l=
 \sum_{i=1}^\infty\lambda_i^{2l}\quad \text{for all}\ l\in\mN\,,\\
 r\equiv\dsp\prod_{i=1}^\infty\frac{1}{1-q_i}=
 \prod_{i=1}^\infty \big(1+\lambda_i^2\big)\leq
 \exp\bigg(\sum_{i=1}^\infty \lambda_i^2\bigg)\,.
\end{gather*}
One shows by induction with respect to $l$ that
\begin{equation}\label{pr-ind}
 \sum_{m_1,m_2,\ldots}\bigg(\sum_{i=1}^\infty
 m_i\bigg)^l\prod_{j=1}^\infty q_j^{m_j}=W_l(p_1,\ldots,p_l)\,r\,,
\end{equation}
where $W_l$ are polynomials. For $l=0$ the l.h.\ side is an
infinite product of geometrical series, so the equality holds with
$W_0=1$. The step from $l$ to $l+1$ is obtained by the application
of the homogeneity operator $\dsp\sum_{i=1}^\infty
q_i\frac{\partial}{\partial q_i}$ to the both sides of the
equation. A~direct calculation yields $\dsp\sum_{i=1}^\infty
q_i\frac{\partial}{\partial q_i}r=p_1r$ and $\dsp\sum_{i=1}^\infty
q_i\frac{\partial}{\partial q_i}p_k=k(p_k+p_{k+1})$, which
confirms the inductive claim and completes the proof of the bound
(\ref{pr-nl}).

\end{document}